\documentclass[a4paper,11pt]{article}
\usepackage[utf8]{inputenc}
\usepackage{amssymb}
\usepackage{amsmath}
\usepackage[english]{babel}
\usepackage{graphicx}
\usepackage{hyperref}
\usepackage{color}
\usepackage{extarrows}
\usepackage[toc,page]{appendix}
\usepackage[a4paper, left=2.5cm, right=2.5cm, top=3.5cm, bott om=3.5cm, headsep=1.2cm]{geometry}
\newcommand{\Lagr}{\mathcal{L}}
\newcommand{\D}{\mathrm{d}}
\usepackage[usenames,dvipsnames]{xcolor}

\allowdisplaybreaks

\begin{document}

\title{Gauge fixing and renormalisation scale independence of tunneling rate in abelian Higgs model and in the Standard Model}
\author{Zygmunt Lalak\footnote{Zygmunt.Lalak@fuw.edu.pl}{}\;  Marek Lewicki\footnote{Marek.Lewicki@fuw.edu.pl}{}\;  and Pawe\l{}  Olszewski\footnote{Pawel.Olszewski@fuw.edu.pl} \\ 
Institute of
Theoretical Physics, Faculty of Physics, University of Warsaw\\ ul. Pasteura 5,
Warsaw, Poland} 
\date{}
\maketitle

\begin{abstract}

We explicitly show perturbative gauge fixing independence of the tunneling rate to a stable radiatively induced vacuum in the abelian Higgs model. We work with a class of $R_\xi$ gauges in the presence of both dimensionless and dimensionful gauge fixing parameters. 
We show that Nielsen identities survive the inclusion of higher order oparators and compute the tunnelling rate to the vacua modified by the nonrenormalisable operators in a gauge invariant manner.
We also discuss implications of this method for the complete Standard Model. 
\end{abstract}

\section{Introduction}
The discovery of the 125 GeV Standard Model Higgs boson, and lack of confirmation of any new physical state in the LHC experiments makes it important to search for possible hints of a possible extension, based on the SM itself.
One of such possibilities is the investigation of the Standard Model  effective potential, which has already been the subject of considerable activity \cite{Degrassi,Ellis:2009tp,Espinosa,Casas:2000mn,Casas:1996aq,Casas:1994qy,Sher,Buttazzo:2013uya}. 

Inclusion of renormalisation group improvement in the Standard Model effective potential reveals an interesting structure at very large  field strengths. Namely a maximum near $10^{11}$ GeV and a new minimum at superplanckian field strength.
The implications depend critically on the value of the measured SM parameters, most importantly the Higgs quartic coupling and the top quark Yukawa coupling. 
For the central value of top and Higgs masses the physical electroweak symmetry breaking minimum turns out to be metastable with respect to the tunnelling to the deeper minimum at superplanckian value of the Higgs field.
This means that the computed lifetime of the SM vacuum is larger than the estimated age of the Universe, and no modification is required to avoid conflict with observations.
However the instability border in the parameter space of $M_{top} - M_{higgs}$ masses, lays uncomfortably close, which means that this result is rather sensitive to modifications brought in by any extension of the SM. 

It is crucial to note, that the whole effect of instability comes from the radiative corrections to the effective action. This means that in the first derivative of the potential with respect to the field, contributions from various radiative corrections cancel against each other and against the tree-level contributions to create additional critical points, not present at tree level. This makes the whole effect particularly interesting and particularly sensitive to new physics, which may appear in the effective action through radiative corrections. A question which becomes relevant in the context of large radiative corrections is the question about gauge independence of observable quantities such as the lifetime of the electroweak vacuum. 

Another important question concerns stability of the SM vacuum after inclusion of ultraviolet completions at or below the Planck scale. This can problem be studied in the spirit of the effective field theory. In \cite{Branchina:2013jra} the neutral Higgs field potential was extended with higher order operators suppressed by suitable powers of the Planck mass.
It was shown that for sensible values of the couplinge electroweak vacuum can be destabilised.
In \cite{Lalak:2014qua} we studied this issue further showing allowed values of the new operators and justifying many approximations used in computation of the lifetime.

Here we study these issues further, from the point of view of the requirement of gauge and scale independence of physical results such as the decay rate. The complete discussion within the SM is rather prohibitive at this point, however, one can learn a lot by studying in detail a simple yet nontrivial example of the abelian Higgs model. Towards the end of this paper we will draw conclusions which can be extended to more general, SM-like models with additional nonrenormalisable operators.

Gauge fixing independence of the observables, i.e. S-matrix elements and physical masses, is in principle a mathematical fact and must hold in any consistent gauge theory. 
This should also be the case for the life-time of a metastable vacuum-like state. 
Given Nielsen identities one can prove that the full effective action $\Gamma [ \Phi ]$ for a configuration of the mean field $\Phi$ which solves the equation of motion $\frac{ \delta \Gamma}{\delta \Phi} =0$ stays invariant with respect to the variation of gauge fixing parameters in covariant gauges, see for instance \cite{Plascencia:2015pga},\cite{Baacke:1999sc}. This is a crucial step and one can accept that this shows that at the formal level the vacuum lifetime is gauge invariant. However, in practice one needs to resort to a perturbative calculation of the effective action and to a quasi-classical determination of the decay rate following classic formulae by Callan and Coleman, and it is a challenge to perform the calculation in such a way that the result stays gauge fixing independent to a given order in the perturbative expansion. In particular, the silent assumption is that to calculate the decay rate one should use the renormalised effective action, which is finite. At the same time the formal proofs 
are usually performed at the level of unrenormalised, formal, expressions, see \cite{Metaxas:1995ab}. 
In this note we shall try to be as explicit as possible in performing the gauge independent calculation at the level of the renormalised abelian Higgs model. We shall consider the renormalised 1-loop effective action to the order $g^6$ in the gauge coupling and check explicitly various assertions and expectations. As usual, we shall be concerned with the exponential dependence of the lifetime on the effective action for the relevant solution of the EOMs. 
In fact, the gauge dependence of the quasi-classical determinat is rather complicated and the proof that it is gauge fixing independent, as it should be, poses a nontrivial challenge, see \cite{Baacke:1999sc}. 
In the proof of gauge fixing invariance the central role is played by Nielsen identities, which read
\begin{equation}
\alpha \frac{\partial \Gamma(\alpha)[\phi]}{\partial \alpha} = \int d^4 x \frac{\delta \Gamma}{\delta \phi_j (x)} H^{\alpha}_j (\phi, x), 
\end{equation}
where $\alpha$ is a gauge fixing parameter. The nonlocal expression $H^{\alpha}_j$ may be expanded in derivatives, see \cite{Metaxas:1995ab},
\begin{equation}
H^{\alpha}_j = C^{\alpha}_j (\phi) + D^{\alpha}_j (\phi) (\partial \phi)^2 + ... \, .
\end{equation}
The coefficients in the expansion of the functional $H_j$ can be computed perturbatively. One possible expansion is in the powers of $\hbar$ - the usual loop expansion. However, in the case of our prime interest, that is in the case of the Standard Model, the nontrivial vacua at large field strength are generated radiatively and their existence results from a delicate balance between the tree level quartic term and higher order corrections. This relies on the approximate 
relation 
$\lambda \sim g^4$,
where $g$ is a gauge coupling or a top Yukawa coupling. This relation should be taken into account when performing the perturbative expansion, since, for instance, there exist contributions of the order $g^6$ at the level of 1 loop and also at the level of 2 loops. Hence, numerically the 2-loop diagrams at the order $g^6$, or higher-loop effects at higher orders could be in principle as important as the lower-loop ones. Of course, this phenomenon depends on a model in question. Firstly, each  additional loop is suppressed by an additional numerical factor of $1/16 \pi^2$. Secondly, the couplings run with energy and the basic relation  $\lambda \sim g^4$ could be modified. 
 
\section{Abelian Higgs model}
Consider a renormalisable $U(1)$ gauge theory of a single scalar matter field. We write the lagrangian, $\Lagr$, for one gauge field $A_\mu$ and two real components of a complex scalar field $\varphi_i$, $i=1,2$.

A two-parameter (quasi) t'Hooft gauge fixing is employed in $\Lagr_{\text{gf}}$. Specifically the dimensionfull parameter $v$ is nonzero and its presence breaks the global gauge symmetry. The $v$ is coupled in $\Lagr_{\text{gf}}$ to the scalar field along the $\varphi_2$ direction, perpendicular to the $\varphi_1$ direction along which we wish to study the quantum corrections. This is no accident and it constitutes a major simplification. The complete Lagrangian takes the form 
\begin{align}
\Lagr &= \Lagr_0 + \Lagr_{\text{int}} + \Lagr_{\text{gf}} + \Lagr_{\psi}\\
\Lagr_0 &= \frac{1}{2} \partial_{\mu} \varphi_i \, \partial^{\mu} \varphi_i -  \frac{1}{2}m^2  \varphi_i \varphi_i - \frac{1}{4} F_{\mu \nu}F^{\mu \nu}  \\
\Lagr_{\text{int}} &= -Z_{\varphi} \, g [ \epsilon_{ij} (\partial_\mu \varphi_i) \varphi_j]A^\mu + Z_{\varphi}\, \frac{g^2}{2} \varphi_i \varphi_i A_\mu A^\mu - Z_{\lambda}\,\frac{\lambda}{4!} (\varphi_i \varphi_i)^2 +\, \Lagr_{\text{ct}} \label{Lagr}\\
\Lagr_{\text{ct}}&=\frac{1}{2} (Z_{\varphi} - 1) \partial_{\mu} \varphi_i \, \partial^{\mu} \varphi_i -  \frac{1}{2}(Z_{m^2} - 1) m^2  \varphi_i \varphi_i - \frac{1}{4} (Z_{A} - 1) F_{\mu \nu}F^{\mu \nu}\\
\Lagr_{\text{gf}} &= - \frac{1}{2 \xi}(\partial_\mu A^\mu + g\,v \varphi_2)^2 \\
\Lagr_{\psi} &= \partial_{\mu} \psi^\ast\, \partial^{\mu} \psi + g\, v \varphi_1\, \psi^\ast \psi \, . \\ 
\end{align}
The expression
$\Lagr_0 + \Lagr_{\text{int}}+ \Lagr_{\text{ct}}$ is invariant with respect to the gauge transformation
\begin{align}
\delta A_\mu &=\partial_\mu \theta \\
\delta \varphi_i &= g\,\theta \epsilon_{ij} \varphi_j \, .
\end{align}

Consequently, the form of the ghost Lagrangian $\Lagr_\psi$ follows from the chosen gauge fixing term:
\begin{equation}
\Lagr_{\psi}= - \psi^\ast \left[ \frac{\delta (\partial_\mu A^\mu + g\,v \varphi_2)}{\delta \theta} \right] \psi = - \psi^\ast \left[ \partial_\mu \partial^\mu -g^2\, v \varphi_1 \right] \psi \, .
\end{equation}

Notice the coupling of ghost fields to $\varphi_1$. Normalisation of the $\theta$ parameter along with the form of $\Lagr_{\text{gf}}$ are chosen such that the kinetic term of $\psi$ is canonical but the gauge coupling $g$ in second power appears in front of the $\varphi_1 \psi^\ast \psi$ term.

The $\xi$ and $v$ are gauge parameters only and as such their values are in principle irrelevant for physical predictions. The fact that one is usually confined to perturbative calculations puts practical limitations on that arbitrariness. Some intermediate results, which, although nonphysical, are rather desirable (for instance Green functions) may not exist in a finite form in a particular renormalisation scheme and for a particular choice of values for $1/\xi$ and $v$. Secondly, since $1/\xi$ and $v$ appear in Feynman rules, one is forced to assign to them an order in the coupling constant which governs the perturbative expansion. For example, in these notes we assume both $\xi$ and $v$ to be formally of order zero in $g$, which is different from $\lambda$ and $m^2$ that we assume to be of order $g^4$. We will yet comment on these issues, as we encounter them later.

Lastly, in case one wishes to shift  the field, $\varphi_1 \rightarrow \left<\varphi \right> + \varphi_1$, it is not an uncommon practice to choose the value of $v$ in such a way that the bilinear terms mixing the $A_\mu$ and $\varphi_2$ fields cancel at the tree level (up to a full derivative). This would correspond to assigning $v=- \xi\left<\varphi \right>$. The constant $\left<\varphi \right>$ could in turn be, say, a minimum of a tree level potential expressed in terms of scalar couplings. We do not follow that rationale in these notes, since we wish to study the $v$-independence and possibly use it as a correctness check.

And yet we do need to shift the field, $\varphi_1 \rightarrow w + \varphi_1$. The reason behind it is that the $w$, aside from being a free, nonphysical parameter, has to get renormalised to counter the loop divergences. The $\Lagr_\text{int}$ above contains all but one counterterms. What is missing, is a divergent $\delta_w$ that we have to add to $w$ after the shift.

\section{Quantum corrections}
We proceed to write down Feynman rules, compute renormalisation group equations and finally the effective action. All of that is done using the background field method: instead of including infinitely many insertions of momentumless legs, one splits the field variable $\varphi_1 = \varphi^\circ + \varphi'$, such that  $\varphi^\circ$ carries the momentumless part of $\varphi_1$. The $\varphi'$ plays the role of propagating quantum field but ultimately vanishes if not hit by a spacetime derivative. The $\varphi^\circ$ on the other hand is treated like a parameter.

Firstly, let's write down the second derivative of the Lagrangian with respect to all fields. That'll give us the inverse of a propagator $D$. (A spacetime derivative hitting a field in the lagrangian translates into mometum in the Feynman rule according to $\partial_\mu \rightarrow - i \, k_\mu$, where $k$ is momentum flowing into the vertex.)
\begin{equation} \label{dwienogi}
\begin{split}
i D^{-1}(k) = diag \Bigg(
k^2-&m^2-\frac{\lambda}{2}{\varphi^\circ}^2 , \;
-(k^2-g^2{\varphi^\circ}^2)\,\left(g^{\mu \nu} - \frac{k^\mu k^\nu}{k^2}\right), \\
&\begin{bmatrix}
      -\left(\frac{k^2}{\xi}-g^2{\varphi^\circ}^2\right) \frac{k^\mu k^\nu}{k^2}, & -i \,k^\nu g\left({\varphi^\circ}+\frac{v}{\xi} \right)   \\[0.6em]
       i \,k^\mu g\left({\varphi^\circ}+\frac{v}{\xi} \right)           , & k^2 - m^2 - \frac{\lambda}{6}{\varphi^\circ}^2 - \frac{g^2v^2}{\xi} \\[0.6em]
\end{bmatrix},\;
k^2 + g v{\varphi^\circ}
\Bigg)\; .
\end{split}
\end{equation}

The diagonal elements correspond to subsequent field subspaces: $\varphi_1$, transverse component of the vector field $A_\mu^T=(g_{\mu \nu} - \frac{k_\mu k_\nu}{k^2})A^\nu$, longitudinal component $A_\mu^L=\frac{k_\mu k_\nu}{k^2}A^\nu$ mixed with $\varphi_2$ and lastly the ghosts $\psi$.
The inverse gives us the propagators
\begin{equation}
\begin{split}
-i D(k) = d&iag \Bigg(
\frac{1}{k^2-m^2 -\frac{\lambda}{2}{\varphi^\circ}^2}, \;
 -\frac{1}{k^2-g^2{\varphi^\circ}^2}\left(g^{\mu \nu} - \frac{k^\mu k^\nu}{k^2}\right), \\
&\frac{1}{\text{D}_N}
\begin{bmatrix}
      [-\xi(k^2-m^2-\frac{\lambda}{6}{\varphi^\circ}^2)+g^2v^2] \frac{k^\mu k^\nu}{k^2}, & -i\, k^\nu g \left( \xi \varphi^\circ + v\right)\\[0.6em]
       i\, k^\mu g\left(\xi \varphi^\circ + v\right)   , & k^2 - \xi g^2{\varphi^\circ}^2           \\[0.6em]
\end{bmatrix},\;
\frac{1}{k^2+g v \varphi^\circ} \Bigg)\; ,
\end{split}
\end{equation}
\begin{gather}
\begin{split}
\text{D}_N = k^4 - k^2(m^2+\frac{\lambda}{6}{\varphi^\circ}^2 - 2 g^2 \varphi^\circ v) + g^2 {\varphi^\circ}^2 \left[\xi\left( m^2 + \frac{\lambda}{6}{\varphi^\circ}^2\right)+ g^2v^2\right] 
=: (k^2-m_+^2)(k^2-m_-^2) \\
m_+^2({\varphi^\circ}^2)=\frac{1}{2}\left(m^2+\frac{\lambda}{6} {\varphi^\circ}^2 \right)  - g^2 v \varphi^\circ +
 \sqrt{\left(m^2+\frac{\lambda}{6} {\varphi^\circ}^2 \right) \left( \frac{1}{4}\left(m^2+\frac{\lambda}{6} {\varphi^\circ}^2 \right) - g^2v \varphi^\circ - \xi g^2 {\varphi^\circ}^2\right)}\\
 m_-^2({\varphi^\circ}^2)=\frac{1}{2}\left(m^2+\frac{\lambda}{6} {\varphi^\circ}^2 \right)  - g^2 v \varphi^\circ -
 \sqrt{\left(m^2+\frac{\lambda}{6} {\varphi^\circ}^2 \right) \left( \frac{1}{4}\left(m^2+\frac{\lambda}{6} {\varphi^\circ}^2 \right) - g^2v \varphi^\circ - \xi g^2 {\varphi^\circ}^2\right)}
\end{split}
\end{gather}
Vertices are straightforwardly read off from the Lagrangian. Resulting Feynman rules are summarised in the Appendix.  

\subsection{Counterterms}
The counterterms which are both necessary and sufficient to cancel all the 1-loop divergencies are:
\begin{align}\label{counter}
\begin{aligned}
Z_A\;&=\;1-\frac{g^2}{24 \pi^2} \frac{1}{\epsilon}\\
Z_\varphi\;&=\;1+\frac{g^2}{8 \pi^2} (3-\xi)\, \frac{1}{\epsilon}\\
Z_w\;&=\;1-\frac{g^2}{8 \pi^2} \frac{v}{w} \,\frac{1}{\epsilon}
\end{aligned}
\hspace{2.5cm}
\begin{aligned}[l]
Z_\lambda \;&=\; 1+\frac{1}{4 \pi^2}\left(9\,\frac{g^4}{\lambda } - g^2 \xi\right)\frac{1}{\epsilon}\\
Z_{m^2}\;&=\;1-\frac{g^2 \, \xi}{8 \pi^2} \, \frac{1}{\epsilon}
\end{aligned}
\end{align}

In particular we do not need additional counterterms in the form of interactions present in the gauge fixing Lagrangian. There is no $\delta_{\Lagr_{\text{gf}}}$. That does not mean that the gauge fixing parameters do not get renormalised. On the contrary, since the renormalisation of the kinetic terms already forced us to define bare gauge coupling and fields: $g_B = \mu^{\epsilon/2}Z_A^{-1/2}g$, $(\varphi_i)_B = Z_\varphi^{1/2} \varphi_i$ and $(A_\mu)_B=Z_A^{1/2} A_\mu$, we would like to have those in the $\Lagr_{\text{gf}}$ as well.

\begin{gather}
\begin{split}
&\Lagr_{\text{gf}} = - \frac{1}{2 \xi}\left(\partial_\mu A^\mu + g\,v \varphi_2\right)^2 =\\\;&= - \frac{1}{2\, Z_A^{1/2} \xi}\left[Z_A^{1/2}\,\partial_\mu A^\mu +\left(\mu^{\epsilon/2}Z_A^{-1/2}g\right)\left(\mu^{-\epsilon/2}Z_A Z_\varphi^{-1/2} v\right) \,Z_\varphi^{1/2} \varphi_2\right]^2 = - \frac{1}{2 \xi_B} \left(\partial_\mu A^\mu_B + g_B\, v_B {\varphi_2}_B\right)^2
\end{split}
\end{gather}

We see that tree-level $\xi$ and $v$ are not the bare, renormalisation scale invariant quantities. We have rather

\begin{gather}
\xi_B = Z_A^{1/2}\, \xi \hspace{0.3cm},\hspace{0.5cm}
v_B = \mu^{-\epsilon/2}Z_A Z_\varphi^{-1/2}\, v \hspace{0.2cm}.
\end{gather}

\subsection{Beta functions}
Given the counterterms from the previous section, one obtains the following 1-loop beta functions:
\begin{align}\label{RGEs}
\begin{aligned}[l]
\beta_g&= \frac{g^3}{48 \pi^2} \\
\gamma_\varphi  &=  -\frac{g^2(3-\xi)}{16 \pi^2} \\
\beta_\lambda &=\frac{3g^2}{4 \pi^2} \left(3 g^2 - \lambda\right) \\
\beta_{m^2} &=  - \frac{3 g^2}{8 \pi^2}m^2 
\end{aligned}
\hspace{2.5cm}
\begin{aligned}[l]
\beta_\xi &= - \xi\frac{g^2}{48 \pi^2} \\
\beta_v &= -v \frac{g^2}{16 \pi^2} \left( \frac{2}{3} + (3-\xi) \right) \\
\beta_w &=\frac{g^2}{16 \pi^2} [ w(3-\xi) - 2 v ] \, .
\end{aligned}
\end{align}
$\beta_g$ warns of the standard Landau pole. Reassuringly the beta functions of $g$, $m^2$ and $\lambda$ do not depend on $\xi$ or $v$. Normally one would also have a 1-loop contribution to $\beta_\lambda$ that is proportional to $\lambda^2$. We neglected it as a contribution of the order $g^8$.

\subsection{Effective action as a sum of diagrams}\label{EffAct}

We wish to calculate a perturbative approximation to the full effective action, usually denoted by $\Gamma[\phi]$, a functional that generates 1PI Green functions. To this end we represent $\Gamma$ as an spacetime integral of an effective Lagrangian, $\Lagr_{\text{eff}}$. The $\Lagr_{\text{eff}}(\phi, \partial_\mu \phi)$ is constructed as a function of fields and their spacetime derivatives, such that the $n$'th functional derivative of its Fourier transform $\;\displaystyle \frac{\delta^n \widetilde{\Lagr}_{\text{eff}}}{\delta^n \widetilde{\phi}}$ is equal to a sum of n-legged 1PI diagrams. Also, we limit ourselves and ask only about the dependence on the $\varphi_1$ field, $\Gamma[\varphi_1]=\Gamma[\varphi_1, \varphi_2\!=\!0, A_\mu\!=\!0,\psi\!=\!0]$.

The computation of $\Lagr_\text{eff}$ hinges on a fact that we employ yet another  expansion. Namely, from all diagramatic contributions we will drop any dependence on the external momenta above the first nontrivial order, i.e. the second power. In other words the (already perturbative in couplings) effective action will be a function of a field variable understood as the value homogeneously filling the four-dimensional spacetime where any position dependence is treated as a perturbation. %

\subsection{Effective potential at 1 loop}

Sum of the vacuum diagrams constructed with the shifted Feynman rules constitutes the momentum independent part of the effective lagrangian, called effective potantial, $V_{\text{eff}}(\varphi^\circ)$. At the level of one loop, every contribution to $V_{\text{eff}}$ is just a closed line of subsequent propagators and masses (two-field couplings). Summing such contributions one readily obtains the well known formula for the so called Coleman-Weinberg one-loop effective potential,
\begin{equation}
V_{\text{eff}}(\varphi^\circ) = -\frac{i}{2} \sum_{\substack{\text{fields:}\\\text{bos } (+1)\\\text{fer } (-2)}}\,\int_k \log \det \left[ iD^{-1}_{\text{field}}(\varphi^\circ,k) \right] \;.
\end{equation}
The matrix under the determinant is exactly the one given in (\ref{dwienogi}).
The sum symbol reminds us to multiply the contribution from ghosts by $-2$ (owing to them being complex scalars that follow the Fermi statistics). We get ($D=4-\epsilon$) 
\begin{gather}\label{CWPot}
\begin{split}
V_{\text{eff}}^{\text{1-loop}}(\varphi^\circ)= - \frac{i}{2}\, \mu^{\epsilon}\! \int\! \frac{\text{d}^D k}{(2 \pi)^D} \bigg[ \log[k^2-m^2-\frac{\lambda}{2} {\varphi^\circ}^2 ] + (D-1) \log[k^2-g^2 {\varphi^\circ}^2] +& \\
+ \log[\text{D}_N] - 2\log[k^2+g^2 v&\varphi^\circ] \bigg]\;.
\end{split}
\end{gather}

Already at this point we can count the powers of $g$.
\begin{equation}
V_{\text{eff}} = V_{g^4}+V_{g^6}+\mathcal{O}(g^8)
\end{equation}

 Since $m^2$ and $\lambda$ are assumed to be of order $g^4$, the first $\log$ in \eqref{CWPot} is of order $g^8$ and we will skip it. The second $\log$ is a nice (gauge-fixing independent) contribution of order $g^4$, which makes it as important as the tree level lagrangian. That is of course part of the design underlying our hope to manufacture qualitative changes in the action via quantum corrections. Using
\begin{equation} \label{ColemanWeinberg}
- \frac{i}{2}\, \mu^{\epsilon}\! \int\! \frac{\text{d}^D k}{(2 \pi)^D} \,\log(k^2-\Delta) = \frac{1}{4} \frac{\Delta^2}{(4 \pi)^2} \left( \log \frac{\Delta}{\bar{\mu}^2} - \frac{3}{2} - \frac{2}{\epsilon} \right)
\end{equation}
one arrives at
\begin{gather}
\begin{split}\label{V4}
V_{g^4}(\varphi^\circ)&= \frac{m^2}{2} {\varphi^\circ}^2 + {Z_\lambda}_{g^0}\frac{\lambda}{4!}{\varphi^\circ}^4 + \frac{1}{4} \frac{(g^2{\varphi^\circ}^2)^2}{(4 \pi)^2} \left[ 3\left( \log \frac{g^2{\varphi^\circ}^2}{\bar{\mu}^2} - \frac{3}{2}- \frac{2}{\epsilon}\right) + \epsilon \,\frac{2}{\epsilon} \right] =\\
&= \frac{m^2}{2} {\varphi^\circ}^2 + \frac{\lambda}{4!}{\varphi^\circ}^4 + \frac{3\,g^4{\varphi^\circ}^4}{64\pi^2} \left( \log \frac{g^2{\varphi^\circ}^2}{\bar{\mu}^2} - \frac{5}{6}\right) + \frac{1}{\epsilon}\left( \frac{(Z_\lambda-1)_{g^0}}{4!}\lambda - \frac{3\,g^4}{32 \pi^2}\right){\varphi^\circ}^4 \, .
\end{split}
\end{gather}

The divergent parts cancel. It is not hard to convince oneself that no multiple-loop diagram contributes at order $g^4$.

The last two logarithms in \eqref{CWPot} should be combined:
\begin{gather}
 \log[\text{D}_N]- 2\log[k^2+g^2 v\varphi^\circ]  = 
 = \log\left[1-\frac{k^2-g^2{\varphi^\circ}^2\,\xi}{(k^2+g^2v \varphi^\circ)^2} \left(m^2+\frac{\lambda}{6}{\varphi^\circ}^2 \right) \right] 
\end{gather}

With the help of dimensional analysis, one can easily see that, after the integration over $k$, only the first term in expansion of the logarithm contributes at order $g^6$.

\begin{gather}\label{22loop1}
\delta V_{g^6}(\varphi^\circ)=- \frac{i}{2}\, \mu^{\epsilon}\! \int\! \frac{\text{d}^D k}{(2 \pi)^D} \left(-\frac{k^2-g^2{\varphi^\circ}^2\,\xi}{(k^2+g^2v \varphi^\circ)^2} \right)\,{m^2}^{(2,2)}({\varphi^\circ}) =i\; \parbox[c]{3.3em}{\includegraphics[scale=0.15]{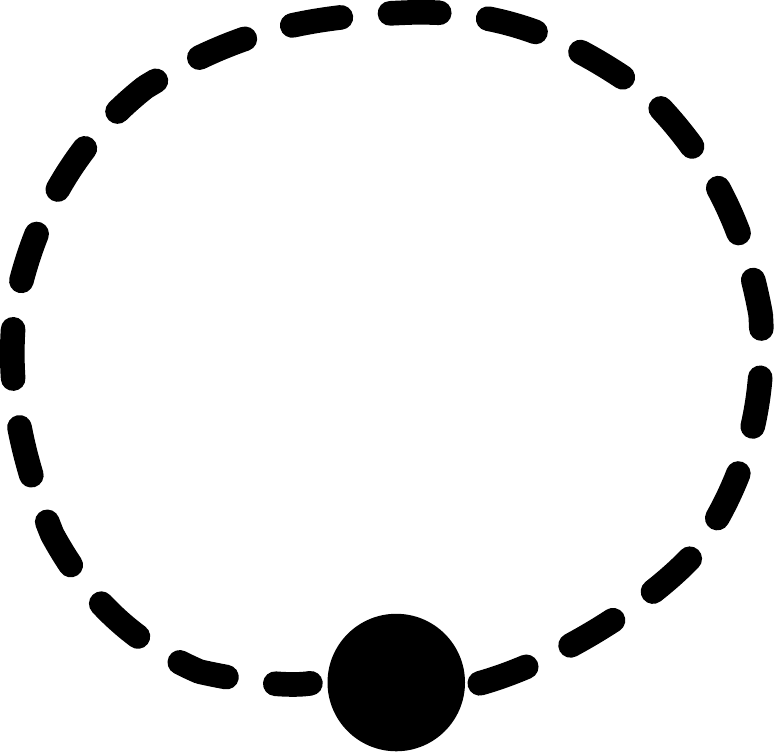}} + \mathcal{O}(g^8),\\ \label{effmass22}
\text{where}\qquad{m^2}^{(2,2)}({\varphi^\circ}) = m^2+\frac{\lambda}{6}{\varphi^\circ}^2 = \frac{1}{\varphi^\circ} \frac{\partial}{\partial \varphi^\circ} V^\text{tree}_{g^4}\;.
\end{gather}

Our $\mathcal{O}(g^6)$ contribution is simply a single $\varphi_2$ propagator, contracted with the tree level selfcoupling of $\varphi_2$, and computed at the lowest order in $g$.

Any remaining $g^6$ correction would originate exclusively from two (and more) loop diagrams. Many of those were analysed in \cite{Metaxas:1995ab}. In this paper, we decide to omit higher loop diagrams. Numerically this is fully justified thanks to the suppression by $(4 \pi)^2$. Also we do not expect them to bring any qualitative novelty to the presented results. 

\begin{figure}[ht]
\begin{center}
  \begin{minipage}[c]{6cm}
    \caption{
       The painlessly included two loop diagrams.
    } \label{twoloop}
  \end{minipage}
  \hspace{0.8cm}  
  \begin{minipage}[c]{4cm}
    \includegraphics[scale=0.32]{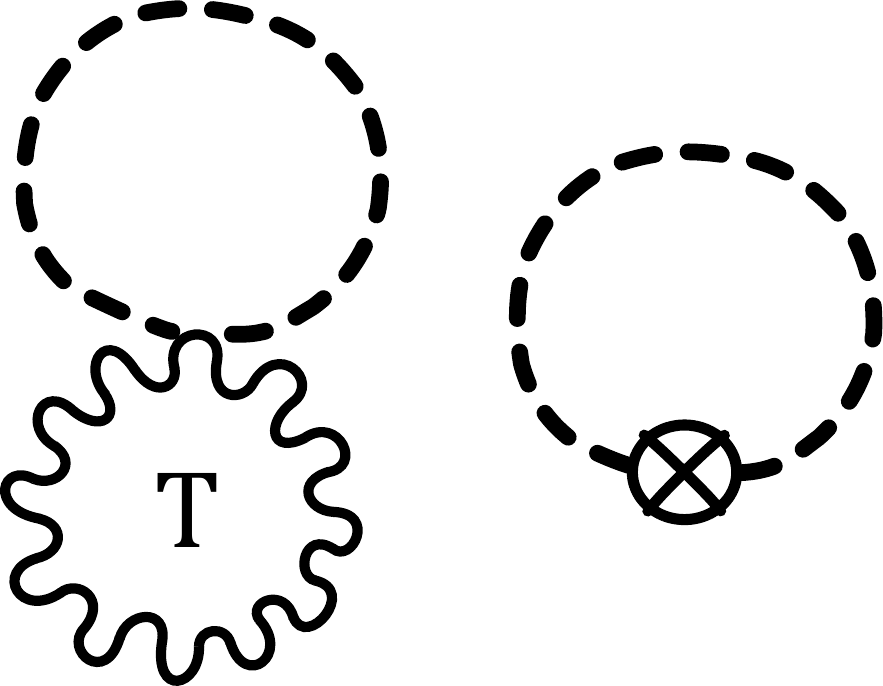}
  \end{minipage}
   \end{center}
\end{figure}

One remark about a particular two loop contribution is in place: One could easily include in his potential the diagrams in Fig. \ref{twoloop} by extending the definition of the effective $\varphi_2$ coupling in \eqref{effmass22} by $\displaystyle {m^2}^{(2,2)} \!\!\rightarrow \widetilde{m^2}^{(2,2)} = \frac{\partial}{\partial \varphi^\circ} V_{g^4}$. Formally the difference between ${m^2}^{(2,2)}$ and $\widetilde{m^2}^{(2,2)}$ in \eqref{22loop1} is, as a two loop correction, beyond our level of accuracy. But we will continue to omit the 'tree' superscript for the sake of convenience.

Performing the integration in \eqref{22loop1} and including all our counterterms  we arrive at
\begin{equation} \label{Vg6}
V_{g^6} = \frac{g^2}{32 \pi^2} \left[v - (2 v + \xi \varphi^\circ) \log \frac{-g^2 v \varphi^\circ}{\bar{\mu}^2}\right] \frac{\partial V_{g^4}(\varphi^\circ)}{\partial \varphi^\circ}\;.
\end{equation}

\subsection{Correction to the kinetic term}

We move on to the calculation of corrections to the kinetic term of the $\varphi_1$ field. To that end, as advertised at the beginning of this chapter, we will use the $\varphi^\circ$ dependent Feynman rules given in the Appendix to calculate the loop corrected $\varphi'$ two-point function, $\left(D^{-1}\right)(p^2)$. Its derivative with respect to $p^2$ is then the desired kinetic term, $\left(D^{-1}\right)'(p^2)= K=K(\varphi^\circ)$,
\begin{equation}
\Lagr_k(\varphi_1) = \frac{1}{2} K(\varphi_1)\; \partial_\mu \varphi_1 \partial^\mu \varphi_1 \;,\quad K = Z_\varphi + \delta K = 1 + K_{g^2}\;. 
\end{equation}
where (cf. also \cite{Metaxas:1995ab})
\begin{equation}
K_{g^2} = \; 3 \frac{g^2}{(4 \pi)^2} \log \frac{g^2 {\varphi^\circ}^2}{\bar{\mu}^2} - \xi \frac{g^2}{(4 \pi)^2}\left( \log\frac{-g^2 v \varphi^\circ}{\bar{\mu}^2} + 1\right) - 2\frac{g^2}{(4 \pi)^2} \frac{v}{\varphi^\circ} \;+\;\mathcal{O}(g^4) \, .
\end{equation}
It is not obvious, at which power of $g$ one should truncate the expansion of $K$ to be consistent with expansion of  the potential up to $g^6$. Of course intuitively, since we have included corrections one order in $g^2$ higher than the tree level potential, we shall do the same for the $K$ function. Another argument for the consistency of such a truncation will be given below in the form of the formulae \eqref{pertNielsen},\eqref{pertNielsen2}.

\section{Renormalisation scale (in)dependence of the action}
We will now solve the RGEs \eqref{RGEs}.
\begin{gather}\label{pertRGE}
\begin{split}
g(\mu) &= g_0 + \frac{g_0^3}{48 \pi^2} \log\!\frac{\mu}{\mu_0} \,+\,\mathcal{O}(g^5)\\
\lambda(\mu) &= \lambda_0 + \frac{3}{4 \pi^2} \left[ \left(3 g_0^4 - \lambda_0 g_0^2 \right) \log\! \frac{\mu}{\mu_0} - \frac{g_0^6} {\pi^2} \log^2\! \frac{\mu}{\mu_0} \right]  \,+\,\mathcal{O}(g^8)\\
m^2(\mu) &= m_0^2 \left(1 - \frac{3\,g_0^2}{8 \pi^2} \log\!\frac{\mu}{\mu_0} \right) \,+\,\mathcal{O}(g^8) \\
\Gamma(\mu) &= \Gamma_0 \left(1 + \frac{g_0^2}{16 \pi^2} (3 - \xi) \log \!\frac{\mu}{\mu_0} \right)\,+\,\mathcal{O}(g^4) \\
\xi(\mu) &= \xi_0 \,+\,\mathcal{O}(g^2) \\
v(\mu) &= v_0 \,+\,\mathcal{O}(g^2) \\
w(\mu) &= w_0 + \frac{g_0^2}{16 \pi^2} \left(w_0(3 - \xi_0 ) - 2v_0\right)\log\! \frac{\mu}{\mu_0} \,+\,\mathcal{O}(g^4) 
\end{split}
\end{gather}
where $g_0 = g(\mu_0)$, etc.

The level of expansion in $g$ is limited by the accuracy of our counterterms \eqref{counter} which was $g^2$. Altogether we are able to write down the effective action including corrections computed at 1-loop up to the order $g^6$ in the coupling constant and up to first power of momentum squared, that is up to  $p^2$.

We can now take our perturbative effective action with its explicit $\mu$ dependence,
\begin{gather}
\begin{split}
\Lagr = \frac{1}{2}\left[1+K_{g^2}(\varphi_1,\, \mu) \right] (\partial \varphi_1)^2 \;-\; V_{g^4}(\varphi_1, \,\mu) - V_{g^6}(\varphi_1, \,\mu) 
\end{split}
\end{gather}
 and plug the running parametrs \eqref{pertRGE} together with the running and shifted field variable in:
\begin{gather}
g \rightarrow g(\mu)\, ,\; ...\\
\varphi_1 \rightarrow \Gamma(\mu) \varphi_1 + w(\mu)\;.
\end{gather}
The $\mu$ dependence from running perfectly cancels the explicit one.  The whole effect on the action amounts to adding the "$0$" subscripts to all parameters and to the renormalisation scale. The $\mu$ parameter disappears completely at the employed accuracy. Taking $\hat{\varphi}(x) = \Gamma_0\, \varphi_1(x) + w_0$, one finds:
\begin{gather} \label{pertAction}
\begin{split}
\quad \Lagr &= \frac{1}{2}\left[ 1+ K_{g^2}(\hat{\varphi}) \right]\partial_\mu \hat{\varphi} \partial^{\mu} \hat{\varphi} - (V_{g^4} + V_{g^6})(\hat{\varphi})\\
K_{g^2} &= 3 \frac{g_0^2}{(4 \pi)^2} \log \frac{g_0^2 \hat{\varphi}^2}{\bar{\mu}_0^2} - \xi_0 \frac{g_0^2}{(4 \pi)^2}\left( \log\frac{-g_0^2 v_0 \hat{\varphi}}{\bar{\mu}_0^2} + 1\right) - 2\frac{g_0^2}{(4 \pi)^2} \frac{v_0}{\hat{\varphi}}\\
V_{g^4} &= \frac{m_0^2}{2} \hat{\varphi}^2 + \frac{\lambda_0}{4!} \hat{\varphi}^4 + \frac{3\,g_0^4 \hat{\varphi}^4 }{64\pi^2} \left( \log \frac{g_0^2 \hat{\varphi}^2}{\bar{\mu}_0^2} - \frac{5}{6}\right) \\
V_{g^6} &= \frac{g_0^2}{32 \pi^2} \left[v_0 - (2 v_0 + \xi_0 \hat{\varphi}) \log \frac{-g_0^2 v_0 \hat{\varphi}}{\bar{\mu}_0^2}\right]\, \frac{\partial V_{g^4}(\hat{\varphi})}{\partial \hat{\varphi}} \, .
\end{split}
\end{gather}

We note that the $\Gamma_0$ and $w_0$ parameters are irrelevant, as they serve only to linearly reparametrize the field variable and disappear when the action is expressed in terms of $\hat{\varphi}$.

\section{Gauge fixing (in)dependence of the action}

As is well known, although the effective action exhibits dependence on the gauge fixing parameters, all physical "observables", when derived consistently in frames of some perturbative expansion, should end up being gauge independent at the level of employed accuracy.
\begin{equation}
\Gamma[\phi] = \Gamma (\xi, v) [\phi]
\end{equation} 
An "observable" could be for instance the value of action computed at some solution to the equation of motion. Schematically
\begin{gather}
\begin{split}
\text{EOM:}& \quad \frac{\delta \Gamma[\phi]}{\delta \phi} \bigg|_{\phi = \phi_{sol}} = 0\;,\\
\text{gauge fixing independence:}& \quad \xi \frac{\partial}{\partial \xi} \Gamma[\phi_{sol}]\,=\, v \frac{\partial}{\partial v} \Gamma[\phi_{sol}] \,=\, 0
\end{split}\label{gfind}
\end{gather}
The least contrived example of such a solution would probably be a homogeneous field value extremizing the effective potential,
\begin{gather}
\varphi(x) = w = const \; , \quad \Gamma[\varphi(x)] = - V_{eff}(w) \int \text{d}^4 x \;,\\
\quad \frac{\partial}{\partial w} V_{eff}(w) = \xi \frac{\partial}{\partial \xi} V_{eff}(\xi,v;w) = v \frac{\partial}{\partial v} V_{eff}(\xi,v;w) =0\;,
\end{gather}
where the gauge invariant is $V_{eff}(w)$.

\subsection{Nielsen identities and vacuum decay}
Another interesting example of an "observable" would be the so-called vacuum decay rate. Assuming there are two nondegenerate minima in $V_{eff}$ and the homogenous field configuration resides in the energetically less favorable one, there generally exists a nonzero chance of tunneling between the minima \cite{Coleman,Callan}. The important point here is that a crucial quantity for determining the tunnelling rate is the action value $S_B$ of a specific solution $\varphi_B$ of the equation of motion. The action functional is derived from the Wick-rotated version of $\Gamma$ (denoted by $\Gamma_E$, signifying that the spacetime metric became Euclidean).
\begin{gather}
\quad S_B = \Gamma_E[\varphi_B] \;,\qquad  \frac{\delta \Gamma_E [ \phi]}{\delta \phi} \bigg|_{\phi = \varphi_B} =0 \quad\text{(+ specific boundary conditions for $\phi_B$)} \\
\xi \frac{\partial }{\partial \xi} S_B = v \frac{\partial }{\partial v} S_B = 0
\end{gather}
It was conjectured that gauge fixing independence of $S_B$ is necessary for arguing that the full tunneling rate is gauge fixing independent  \cite{Metaxas:1995ab}.

Looking at \eqref{gfind}, we see that there should exist a functional $H[\phi]$ such that
\begin{equation}
\alpha \frac{\partial \Gamma[\phi]}{\partial \alpha} = \left. \int \right. H^\alpha [\phi] \frac{\delta \Gamma[\phi]}{\delta \phi} \;,
\end{equation}
where $\alpha$ denotes a generic gauge fixing parameter and $H^{\alpha} = C^{\alpha} (\phi) + D^{\alpha} (\phi) (\partial \phi)^2 + ...$.

The above formula is the famous Nielsen identity. It was formally derived by Nielsen without using a specific form of the action \cite{Nielsen:1975fs}. The generic derivation is made possible by the BRST invariance present in the Lagrangian of any gauge field theory. But notably the finiteness of $H[\phi]$ is not guaranteed.

The Nielsen identities for the abelian Higgs model were carefully rederived and studied in detail in \cite{Aitchison}.

In \cite{Metaxas:1995ab} they were used in the context of the tunneling problem in the same model. The gauge fixing independence was examined there on top of a perturbative expansion in the gauge coupling constant $g$. We now summarize a few points from that work. The Authors showed that the Nielsen identities, after expanding in $g$ and field momentum (in the manner used also in these notes), produce the following identities
\begin{gather} \label{pertNielsen}
\xi \frac{\partial K_{g^2}}{\partial \xi}  = 2 \frac{\partial C_{g^2}^\xi}{\partial \varphi_1} \; , \quad
\xi \frac{\partial V_{g^6}}{\partial \xi}  = C_{g^2}^\xi \frac{\partial V_{g^4}}{\partial \varphi_1} \; ,\\ \label{pertNielsen2}
 v  \frac{\partial K_{g^2}}{\partial  v }  = 2 \frac{\partial C_{g^2}^v }{\partial \varphi_1} \; , \quad
 v  \frac{\partial V_{g^6}}{\partial  v }  = C_{g^2}^v  \frac{\partial V_{g^4}}{\partial \varphi_1} \; .
\end{gather}
We use our expressions \eqref{pertAction} to explicitly check these identities and compute the functions $C^\alpha_{g^2}$:
\begin{gather}
\begin{split} \label{Cfunctions}
C^\xi_{g^2} = - \frac{g_0^2}{32 \pi^2}\,\xi_0 \hat{\varphi} \,\log \!\frac{-g_0^2\, v_0 \hat{\varphi}}{\bar{\mu}_0^2} \; ,\\
C^v_{g^2} = - \frac{g_0^2}{32 \pi^2} \left[\xi_0 \hat{\varphi} +v_0 \left(2\log \!\frac{-g_0^2\, v_0 \hat{\varphi}}{\bar{\mu}_0^2} +1\right) \right] \; .
\end{split}
\end{gather}
(This is not to say that $C$ functions may only be inferred from Nielsen identities. On the contrary, in the context of a perturbative calculation, they have their own representation as a sum of 1PI diagrams \cite{Aitchison}.)

Assumed finiteness of $C$'s allowed the Authors to derive the gauge-fixing independence of $S_B$ in the context of perturbative calculation done up to $g^6$, which is, as we've seen, the lowest non-trivial order exhibiting gauge fixing parameters. Consider first how the order of expansion in $g$ translates into expansion of the solution $\varphi_B$,
\begin{gather}
\Lagr = \Lagr^0 + \Lagr^1 \,+\, ...\quad,\qquad
\Lagr^0 (\varphi)= \frac{1}{2} \left( \partial_\mu \varphi \right)^2 + V_{g^4} (\varphi) 
\\
\varphi_B = \varphi^0_B + \varphi_B^1 \,+\, ...\qquad \text{where by definition,} \label{bouncesolution}
\\
0=\frac{\delta \Lagr^0 (\varphi)}{\delta \varphi} \bigg|_{\varphi = \varphi_B^0} \; \Leftrightarrow\;\partial_\mu^2 \varphi_B^0 = V_{g^4}'(\varphi_B^0)\; .
\end{gather}
And further into the expansion of the action
\begin{gather}
S^0_B = \int \!\text{d}^4 x\;\Lagr^0(\varphi^0_B) \qquad \text{(which is explicitly gauge fixing independent)}   \label{actions}\\
S_B^1 = \int\!\text{d}^4 x \left[ \frac{\delta \Lagr^0(\varphi)}{\delta \varphi} \bigg|_{\varphi = \varphi_B^0} \!\cdot\varphi_B^1 \; +\; \Lagr^1(\varphi^0_B)\right] =  \int\!\text{d}^4 x \;\Lagr^1(\varphi^0_B)\;.
\end{gather}
Now we hit it with the derivative with respect to the gauge fixing parameter(s),
\begin{gather}\label{gaugeSB}
\begin{split}
\alpha & \frac{\partial}{\partial \alpha} \,S_B =  \int\!\text{d}^4 x \;\alpha \frac{\partial}{\partial \alpha} \,\Lagr^1(\varphi_B^0) = \\
&= \int\!\text{d}^4 x \left[ \frac{1}{2} \alpha \frac{\partial}{\partial \alpha} \, K_{g^2}\,(\partial_\mu \varphi)^2 \;+\;  \alpha \frac{\partial}{\partial \alpha} \,V_{g^6} \right] \bigg|_{\varphi=\varphi_B^0} =
 \int\!\text{d}^4 x \left[ \frac{\partial C^\alpha_{g^2} (\varphi)}{\partial \varphi} \,(\partial_\mu \varphi)^2 \;+\; C^\alpha \frac{\partial V_{g^4}}{\partial \varphi} \right] \bigg|_{\varphi = \varphi_B^0} = \\
&=\int\!\text{d}^4 x \left[ \underbrace{\frac{\partial}{\partial \, x^\mu}\left( C^\alpha_{g^2}(\varphi) \,\partial_\mu \varphi \right)\bigg|_{\varphi = \varphi_B}}_{\text{boundary conditions  for $\varphi_B^0$}} \;+\; C^\alpha \underbrace{\left(
- \partial_\mu^2 \varphi \,+\, \frac{\partial}{\partial \varphi}V_{g^4} \right)\bigg|_{\varphi = \varphi_B}}_{\text{EOM}} \right] \; = \; 0\;,
\end{split}
\end{gather}
where we used \eqref{pertNielsen}. We end up with zero, thanks to the reasons specified under the horizontal braces.

This closes the discussion of gauge fixing dependence of the action computed at any solution to the equation of motion in the abelian Higgs model, computed up to the corrections of order $g^6$ at one loop in the perturbative expansion. The dependence cancels, under the assumption that $(\log \varphi) \,\partial_\mu \varphi$ vanishes at the boundaries. Notice also that, at this accuracy, one does not include $\varphi^1_B$ in his computation of $S_B$.

\subsection{Depth of the minima}
For the sake of completeness of our discussion, using the same approach as above, we will examine the corrections to the position and value of the extrema of the effective potential.
\begin{gather}
\varphi_{min} = \varphi_{min}^0 + \varphi_{min}^1 +... \quad; \qquad 
0 = \frac{\partial\,V_{g^4}}{\partial \varphi} \bigg|_{\varphi = \varphi^0_{min}} 
\\
0 = \frac{\partial(V_{g^4} + V_{g^6} + ...)}{\partial \varphi} \bigg|_{\varphi = \varphi_{min}} 
= \frac{\partial^2\,V_{g^4}}{\partial \varphi \,^2} \bigg|_{\varphi = \varphi^0_{min}} \!\! \cdot \varphi^1_{min} \;+\; 
\frac{\partial\,V_{g^6}}{\partial \varphi} \bigg|_{\varphi = \varphi^0_{min}} +...\; = \cdots\\
\text{recall that} \qquad V_{g^6} = f(\varphi) \frac{\partial\,V_{g^4}}{\partial \varphi} \;, \quad
f(\varphi) = \frac{g_0^2}{32 \pi^2} \left[ v_0 - (2 v_0 + \xi_0 \varphi) \log\frac{-g_0^2 v_0 \varphi}{\bar{\mu}_0^2} \right]\;, \\
\frac{\partial\,V_{g^6}}{\partial \varphi} \bigg|_{\varphi = \varphi^0_{min}} \!= f(\varphi_{min}^0) \frac{\partial^2\,V_{g^4}}{\partial \varphi\,^2} \bigg|_{\varphi = \varphi^0_{min}} \;,\quad \text{so that}\\
\cdots = \;\frac{\partial^2\,V_{g^4}}{\partial \varphi\,^2} \bigg|_{\varphi = \varphi^0_{min}}\left(\varphi_{min}^1 + f(\varphi_{min}^0)\right)\;+\mathcal{O}(g^8) \qquad \text{and finally} \\
\varphi^1_{min} = - f(\varphi_{min}^0)
\end{gather} 
\begin{gather}
\begin{split}
V(\varphi_{min}) - V_{g^4}(\varphi_{min}^0) =& V_{g^4}'(\varphi_{min}^0) \!\cdot \!\varphi_{min}^1 + V_{g^6}(\varphi_{min}^0) \,+\mathcal{O}(g^8) =\\
=& \underbrace{V_{g^4}'(\varphi_{min}^0)}_{0} \underbrace{\left( \varphi_{min}^1 + f(\varphi_{min}^0) \right)}_{0} \,+\,\mathcal{O}(g^8)
\end{split} \label{noMinShift}
\end{gather}
At the lowest order, an extremum resides in $\varphi_{min}^0$ with the value of $V_{g^4}(\varphi_{min}^0)$. Including corrections one level of $g^2$ higher has an effect only in non-Landau gauges: It shifts $\varphi_{min}$ by a value of order $g^2$, but the height of the extremum does not change.

Of course, if one were to just start plotting our $V_{g^4} + V_{g^6}$ with different choices of $\xi_0$ and $v_0$, he would observe the minima going up or down by some nonzero value since he would implicitly have not thrown away all of the $\mathcal{O}(g^8)$ from \eqref{noMinShift}.

\subsection{Physical mass of $\varphi_1$}

Physical mass of the $\varphi_1$ field (the pole mass) should also stay independent of the gauge fixing. We would like to demonstrate this in our simple setup.

Physical mass squared, $m^2$, is defined as a pole of the inverse two-point function regarded as a function of the momentum squared,
\begin{equation}\label{pole}
K(\varphi_{min})\, m^2 - V''(\varphi_{min}) = 0\,.
\end{equation}
Value of the field variable used here, $\varphi_1 = \varphi_{min}$, is a minimum of the potential, and corresponds to one of the vacua.
Again, we search for perturbative solution to \eqref{pole}:
\begin{equation}
m^2 = m^2_{g^4} + m^2_{g^6} + \ldots
\end{equation}
\eqref{pole} reads now
\begin{equation}
\left[ \left(1+ K_{g^2} + \ldots \right)\left(m^2_{g^4} + m^2_{g^6} + \ldots \right) = V''_{g^4} + V''_{g^6} + \ldots \right]\bigg|_{\varphi_{min}^0 +  \varphi_{min}^1 + \ldots}
\end{equation}
Hence, at the lowest order we have
\begin{equation}
m^2_{g^4} = V''_{g^4}(\varphi_{min}^0)\;.
\end{equation}

Further, we need to remember, that
\begin{gather}
V_{g^6} = f(\varphi) V'_{g^4}(\varphi)\quad,\qquad K_{g^2}(\varphi) = 3 \frac{g_0^2}{(4 \pi)^2} \log \frac{g_0^2\varphi^2}{\bar{\mu}^2} + 2 f'(\varphi) \\
\varphi_{min} = \varphi_{min}^0 +  \varphi_{min}^1 + \ldots \quad,\qquad V'_{g^4}(\varphi_{min}^0) = 0 \quad,\qquad \varphi_{min}^1 = - f(\varphi_{min}^0)\;,
\end{gather}
and ultimately it is only the function $f$ that depends on gauge fixing parameters. The next contribution to the mass squared is now given by
\begin{gather}
m^2_{g^6} + K_{g^2}(\varphi_{min}^0) \, m^2_{g^4} = V'''_{g^4}(\varphi_{min}^0)\, \varphi_{min}^1 + V''_{g^4}(\varphi_{min}^0) \\
\begin{split}
m^2_{g^6} = - \left[ 3\frac{g_0^2}{(4 \pi)^2} \log \frac{g_0^2 \hat{\varphi}^2}{\bar{\mu}_0^2} + 2 f'(\varphi_{min}^0) \right] V''_{g^4}(\varphi_{min}^0) + V'''_{g^4}(\varphi_{min}^0)\left[ -f(\varphi_{min}^0) \right] \\
+  f''(\varphi_{min}^0) \underbrace{V'_{g^4}(\varphi_{min}^0)}_{0}
+ 2 f'(\varphi_{min}^0) V''_{g^4}(\varphi_{min}^0) + f(\varphi_{min}^0) V'''_{g^4} (\varphi_{min}^0) = \\
 = - 3\frac{g_0^2}{(4 \pi)^2} \log \left(\frac{g_0^2 \hat{\varphi}^2}{\bar{\mu}_0^2}\right) V''_{g^4}(\varphi_{min}^0)
\end{split}
\end{gather}
We observe that $m^2_{g^4}$ and $m^2_{g^6}$ are $\xi$ and $v$ independent. Notably, to reach this conclusion, it is crucial to remember about gauge-fixing dependence of $\varphi_{min}$, the position of a minimum in the potential. The mass depends on $\xi$ and $v$ both via their explicit presence in the action, and implicitly through $\varphi_{min}$. It is again the Nielsen identity, that guarantees that the two dependencies cancel each other.

\subsection{Removing gauge dependence}
Let us take a step back and rethink why do we bother with an unspecified (but only parametrized) gauge fixing and what would be the alternative.

Ending with a gauge dependent formula for something that was supposed to be a physical quantity, raises warning flags. There may be an unrecognised implicit gauge dependence left in some of the variables. Or perhaps one's perturbative calculation wasn't hundred percent consistent all the way through and relies on some estimates. Or maybe the result is plain wrong.
Being gauge independent doesn't of course guarantee that a formula represents a physical quantity, but it is a reasonably optimistic sign.

Following this line of thought, it is most desirable for the gauge dependence to vanish at the very last step of a calculation. It does not need to be so, one may get "luckly" and witness a cancelation at an earlier stage. A good example is the Coleman-Weinberg $U(1)$ model as considerd here, but written in terms of radial coordinates for the scalar field \cite{Tye:1996au},
\begin{gather}
\varphi_1 + i\, \varphi_2 = \rho \, e^{ i\sigma}\,.
\end{gather}
In this approach Feynman rules appear gauge dependent, but the dependence cancels already at the moment of contracting vertices with propagators.

Another moment to walk away from gauge dependence in our model would be at the level of the effective lagrangian \eqref{pertAction}. There one could make a field redefinition to normalize the kinetic term,
\begin{equation}
K  (\partial \,\varphi)^2 =: \left( \partial \, \tilde{\varphi}(\varphi) \right)^2\,,
\end{equation}
invert for $\varphi(\tilde{\varphi})$ and plug it into the potential. The resulting expression for $\Lagr(\tilde{\varphi})$ would be explicitly gauge independent.

The point we wish to make is that there is nothing hard in removing the gauge dependence. The methods mentioned above are rather general. But they are no better than the simplest method of all, which is to fix the gauge at the very beginning.

\section{Nonrenormalizable interactions}

Although we are obviously dealing with a toy model, we would like to draw lessons applicable to more realistic models. 
Those can often be represented as effective field theories and one may need to include nonrenormalisable interactions in their Lagrangian.
Following this reasoning we will discuss addition of simplest nonrenormalisable terms which are most relevant for the effective potential:
\begin{equation}
\delta \Lagr_{g^4} \,=\, \frac{\lambda_6}{6} \frac{ (\varphi_i \varphi_i)^3 }{\Lambda^2}\;+\;
\frac{\lambda_8}{8} \frac{ (\varphi_i \varphi_i)^4 }{\Lambda^4}\;+\,... \; .
\end{equation}
Immediately we had to assign the power in $g$ to the new couplings $\lambda_k$. In order for them to be visible at the level of our previous calculations, they have to be $\mathcal{O}(g^6)$ at the most. However, since we are curious about loop effects introduced with the new interactions, we have no other choice than to assume $\lambda_k = \mathcal{O}(g^4)$ for at least some of the $k = 6,8,...$.

Consequently, we have a new tree level terms in the potential,
\begin{gather}
\delta V_{g^4} \,=\, \frac{\lambda_6}{6} \frac{{\varphi^\circ}^6}{\Lambda^2} \;+\;
\frac{\lambda_8}{8} \frac{{\varphi^\circ}^8}{\Lambda^4}\;+\,...\, ,
\end{gather}
but also new vertices,
\begin{gather}
\sim \;\varphi_1^4 \varphi_2^2 \;,\; \varphi_1^6 \varphi_2^2 \;,\,...
\end{gather}
with the property that a pair of their $\varphi_2$ legs may be contracted and form an $\mathcal{O}(g^2)$ loop. It is easy to quickly reproduce the induced loop corrections, simply updating the effective mass given by $\frac{\partial V_{g^4}(\varphi^\circ)}{\partial \varphi^\circ}$ in \eqref{Vg6}. We get the contribution to $V_{g^6}$  
\begin{gather}\nonumber
\delta^{\text{loop}}\, V_{g^6} = \frac{i}{2} \left[ \frac{i}{8 \pi^2} (-g^2 v \varphi^\circ) \left(\frac{2}{\epsilon} + \frac{1}{2} \!-\! \log \frac{-g^2 v \varphi^\circ}{\bar{\mu}^2} \right) - \frac{i}{16 \pi^2} (g^2 {\varphi^\circ}^2 \xi) \left(\frac{2}{\epsilon} - \log\frac{-g^2 v \varphi^\circ}{\bar{\mu}^2} \right) \right]  \frac{1}{\varphi^\circ} \frac{\partial  \delta V_{g^4}}{\partial \, \varphi^\circ}
\end{gather}
and required counterterms
\begin{gather}
\begin{split}
\delta^{\text{tree}}\, V_{g^6} \, = \,
& \left( Z_{\lambda_6}\! -\! 1 \right)\frac{\lambda_6}{6} \frac{{\varphi^\circ}^6}{\Lambda^2} \, + \, \left( Z_{\lambda_8}\! -\! 1 \right)\frac{\lambda_8}{8} \frac{{\varphi^\circ}^8}{\Lambda^4} \;+\, ...\,+\;
 \left(\lambda_6 \frac{{\varphi^\circ}^5}{\Lambda^2} + \lambda_8 \frac{{\varphi^\circ}^7}{\Lambda^4} \right) \delta_w \;+\,...\, , 
\end{split}
\end{gather}
where
\begin{gather}
Z_{\lambda_k} \;=\;1 - \xi\, \frac{k\,g^2}{16 \pi^2}\, \frac{1}{\epsilon}\;.
\end{gather}
The RGEs as usual may be computed requiring that the renormalised coupling do not depend on $\mu$ or simply extracted from $\delta^{\text{loop}}V_{g^6}$. Both methods result in
\begin{gather}
\beta_{\lambda_k} \;=\; -k \frac{3\,g^2}{16 \pi^2} \lambda_k \\
\lambda_k \; = \; {\lambda_k}_0 \left( 1 - k \frac{3\,g^2}{16 \pi^2} \log\frac{\mu}{\mu_0} \right) + \mathcal{O}(g^8)
\end{gather}
and finally
\begin{gather}
\begin{split}
\delta^{\text{nonren}}\, V \;=\; 
\frac{{\lambda_6}_0}{6} \frac{\hat{\varphi}^6}{\Lambda^2} \;+\;
\frac{{\lambda_8}_0}{8} \frac{\hat{\varphi}^8}{\Lambda^4}\;+\,...\,+\hspace{6cm}&\\
+\;\frac{g_0^2}{32 \pi^2} \left[ v_0 - (2 v_0 + \xi_0 \hat{\varphi}) \log\frac{-g_0^2 v_0 \hat{\varphi}}{\bar{\mu}_0^2} \right] 
\cdot\left( 
{\lambda_6}_0 \frac{\hat{\varphi}^4}{\Lambda^2} \;+\;
{\lambda_8}_0 \frac{\hat{\varphi}^6}{\Lambda^4}\;+\,...\, \right)
\end{split}
\end{gather}

An important point here, is that the above result is completely compatible with and could have been to large extent obtained from the Nielsen identities in \eqref{pertNielsen}, \eqref{pertNielsen2}. This means in particular that the proof of the gauge fixing independence of the bounce's action $S_B$, \eqref{gaugeSB}, goes through unchanged.


\section{Numerical study of the model}

In this section we present the numerically obtained results aimed to illustrate the effective potential and tunneling solutions between its vacua as well as the higher order corrections to these results.

\subsection{Shape of the potential}

Let us begin by reparametrising  the lowest order potential $V_{g^4}$, defining $c_{m^2}$, $c_{\lambda}$ and $x$,
\begin{gather}
m_0^2 =: c_m \, \frac{3 \,g_0^4}{16 \pi^2} \bar{\mu}_0^2 \qquad
\lambda_0 = : c_{\lambda} \,  \frac{3 \,g_0^4}{8 \pi^2} \qquad
x := \frac{\varphi^2}{\bar{\mu}_0^2} \\
V_{g^4} / \bar{\mu}_0^4 = \frac{g_0^4}{64 \pi^2} \left[ 6 c_m x + \left( c_\lambda + 3 \log g_0^2 - \frac{5}{2} + 3 \log x\right)x^2 \right] \; .
\end{gather}
Note that putting $c_m$ and $c_\lambda$ of order one is consistent with our power counting in $g$. That is, assuming that the chosen $\bar{\mu}_0^2$ would be in fact some scale characteristic for the processes one wishes to describe using  the potential.
 
One extremum of $V_{g^4}$ is of course at $\varphi = 0$. Others have to satisfy the equation
\begin{gather}
\frac{1}{\varphi / \bar{\mu}_0 } \frac{ \partial\, V_{g^4} / \bar{\mu}_0^4}{ \partial \, \varphi / \bar{\mu}_0 } = 
\frac{g_0^4}{16 \pi^2} \left[ 3 c_m + x \left(c_\lambda + 3 \log g_0^2 -1  + 3 \log x \right) \right] =: \frac{3\, g_0^4}{16 \pi^2} c_m \, h(x) = 0\; .\\
h(x) = 1 + \frac{x}{A} \left( \log x  - B \right) \quad,\qquad A = c_m \;,\;\;B= \frac{1 \!-\! c_\lambda}{3} - \log g_0^2
\end{gather}
In the two-dimensional space of $A$ and $B$ parameters, there are two interesting regions. One where $h(x)=0$ has no solutions and one with two such solutions  (at the boundary there is one solution). The region with two solutions is given by the constraints
\begin{gather}
0 \,<\, A \,<\, e^{B-1}\quad,\qquad \text{which translates to} \qquad g_0^2 \, <\, \frac{1}{c_m} e^{-\frac{2+c_\lambda}{3}} \quad \wedge \quad c_m > 0
\end{gather}
For example, when $c_m = 1$, varying $c_\lambda$ in range $(-1,1)$ changes the upper bound on $g_0$ respectively in  range $(0.8,0.6)$. Analogously the range $(-\pi^2, \pi^2)$ translates to $(3.7, 0.1)$. It is fair to say, that $\lambda$ and $m^2$ being roughly of order $g^4$ are indeed consistent with radiative symmetry breaking. Also under this assumption, it doesn't matter much, whether $c_\lambda$ is positive or negative. If we were inclined to strain the consistency of the perturbative calculation by going with $c_m$ to higher values, the structure of the minima could be preserved by making either $\log g_0^2$ or $c_\lambda$ negative.

\subsection{Simple case study}

\begin{figure}[ht]
\begin{center}
 \includegraphics[scale=0.32]{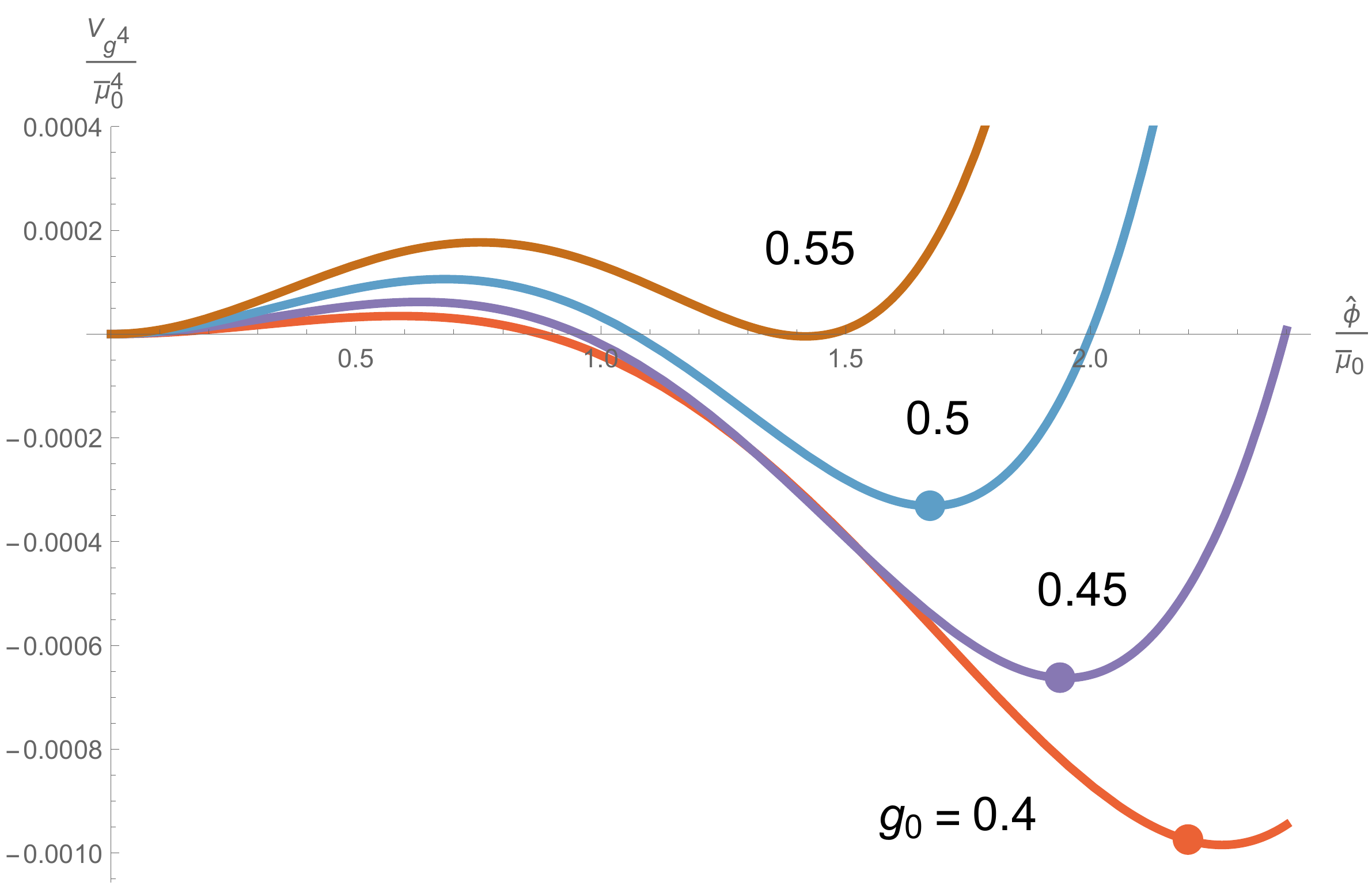}
 \includegraphics[scale=0.37]{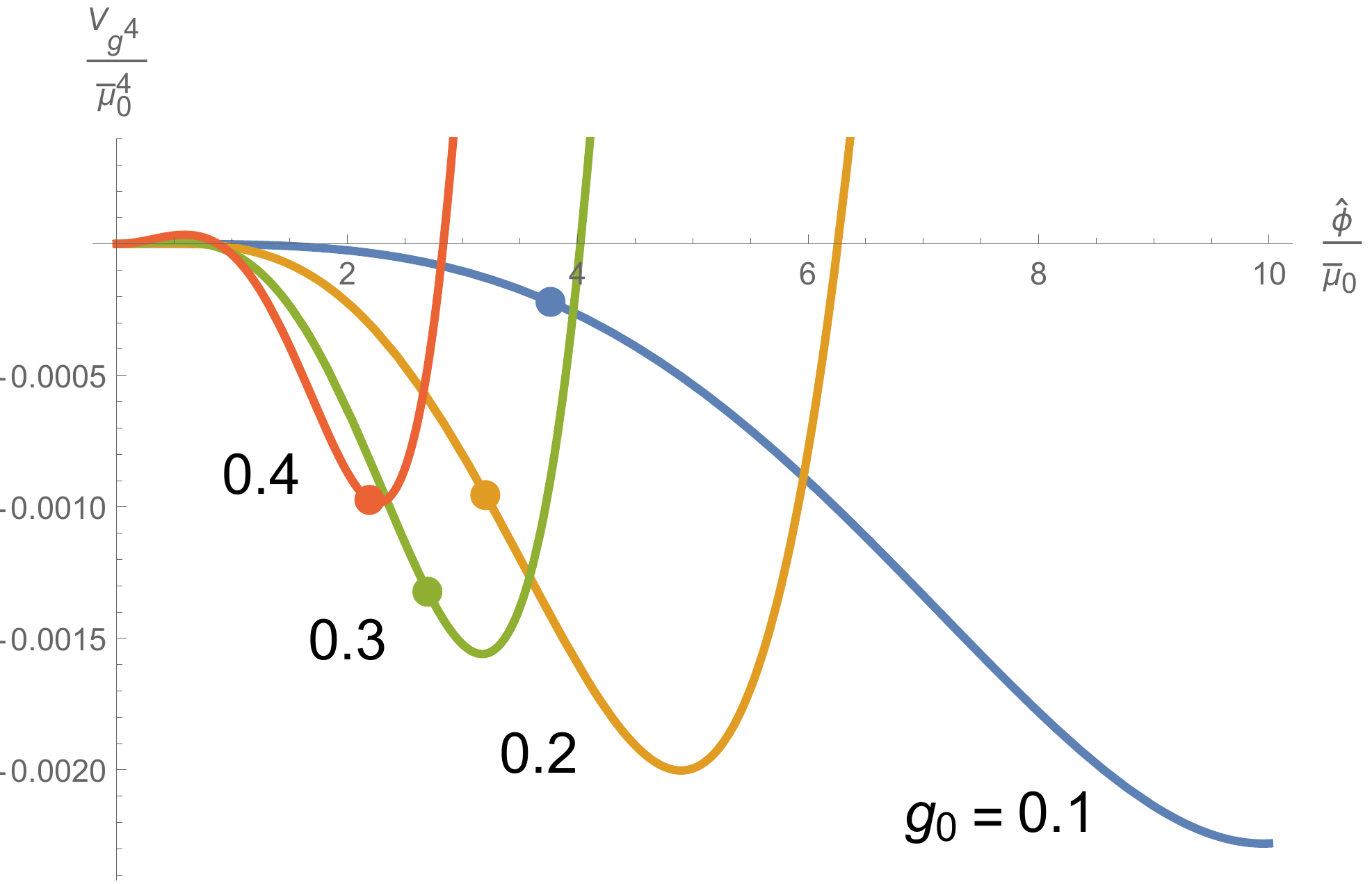}
 \end{center}
 \caption{Plots of the potential at the lowest order, $V_{g^4}$, for a specific choice of couplings (see text). The renormalisation scale $\mu_0$ is used as a unit of energy.}
 \label{potentials}
\end{figure}

Figure \ref{potentials} shows the potential for few values of $g_0$ between $0.1$ and $0.55$ choosing the other parameters equal to one  $(c_\lambda, c_m, g_0)=(1, 1, g_0)$. Note that this explicitly makes both $\lambda_0$ and $m^2_0 \,$ functions of $g_0$.
While $\mu_0$ may be thought of as being substituted with our unit of energy.

We observe that abelian Higgs model allows us to study both cases: when the two minima are nearly degenerate and when the second minimum is much deeper and further from the first one. The second case is more closely associated with the issue of electroweak vacuum instability in the Standard Model.

\begin{figure}[ht]
\begin{center}
  \includegraphics[scale=0.39]{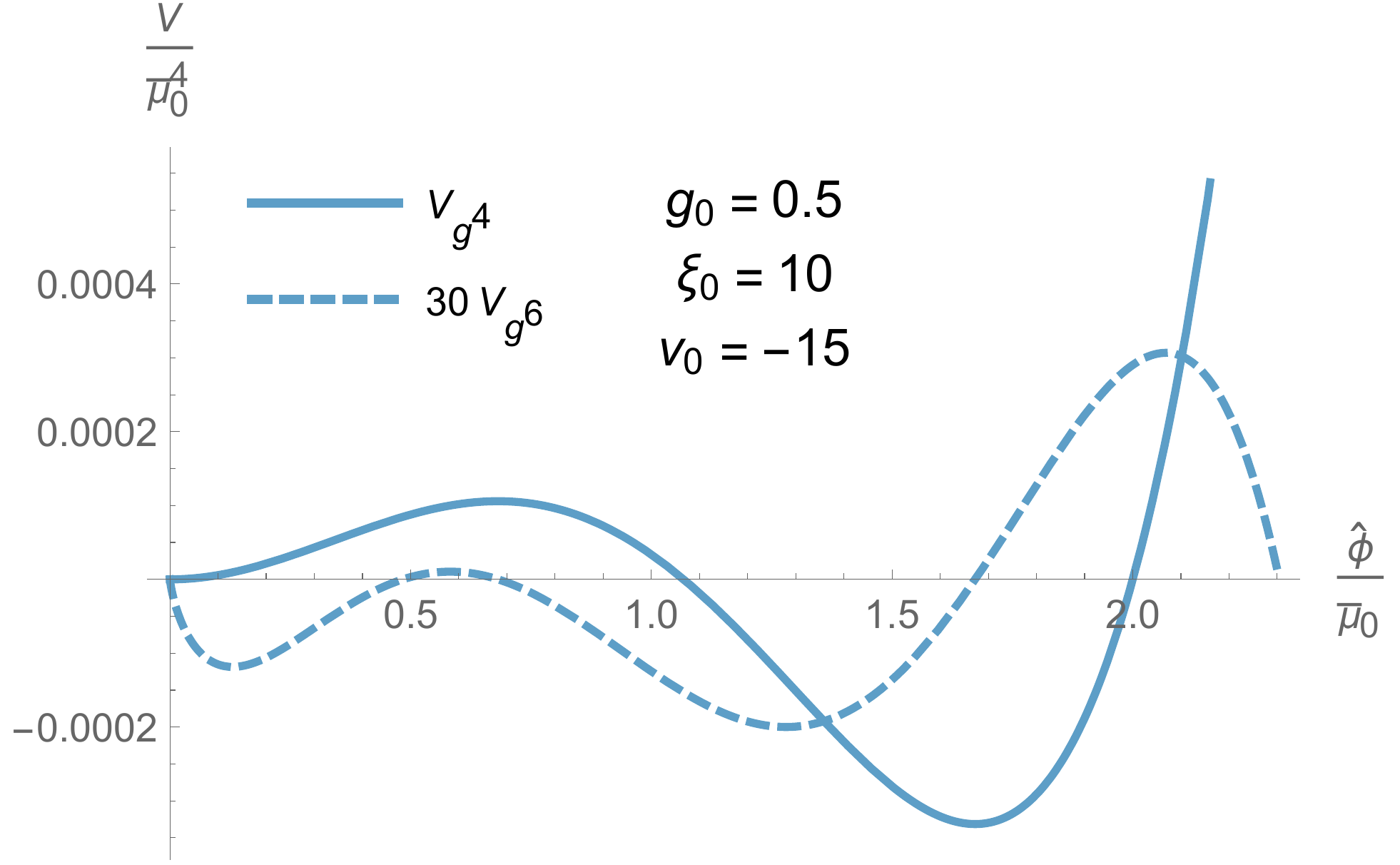}
  \includegraphics[scale=0.42]{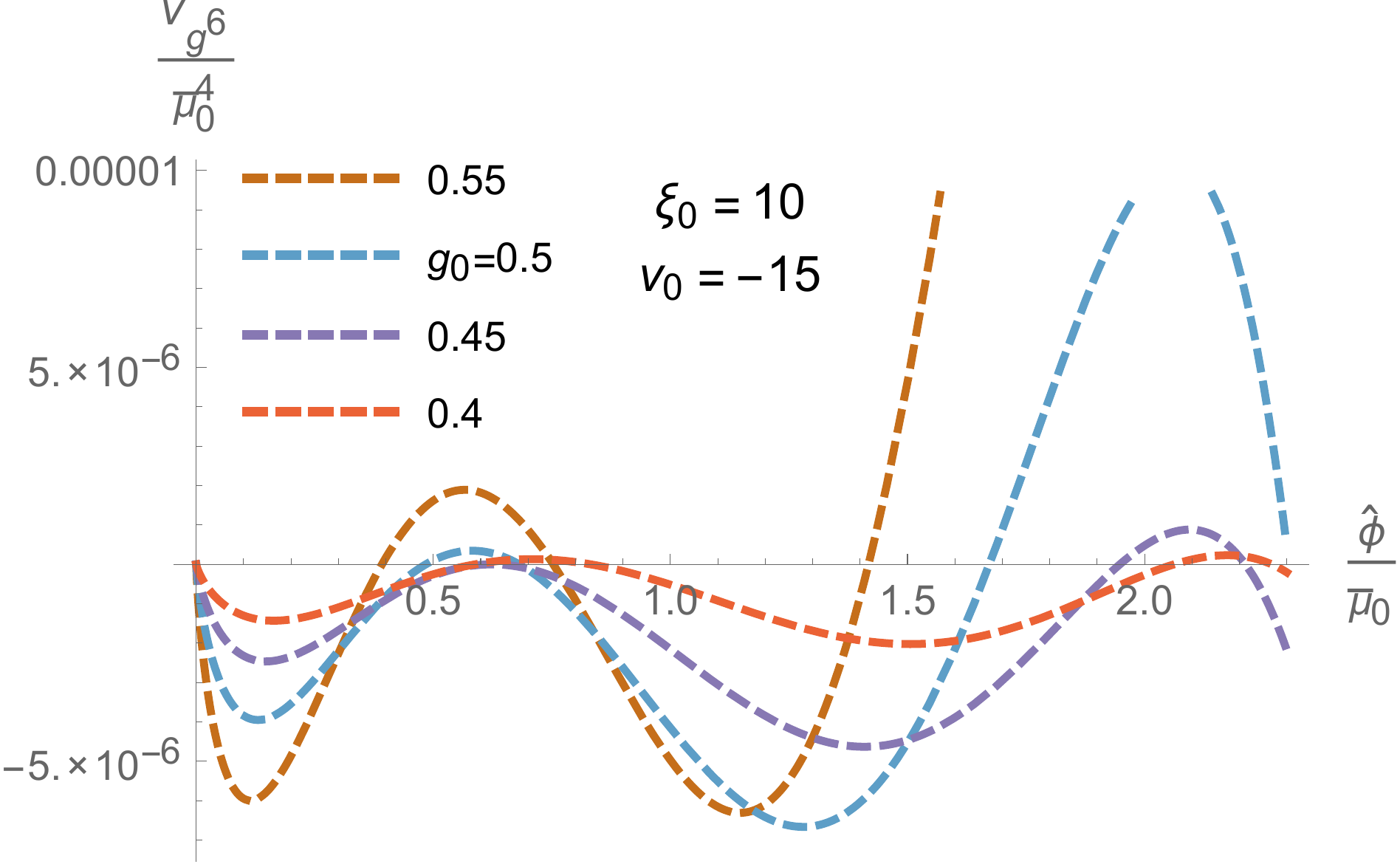}
 \end{center}
 \caption{Left panel: the higher order correction to the potential multiplied by $30$ and plotted along the $V_{g^4}$. Right panel: the correction for different values of $g_0$.}
 \label{pot3}
\end{figure}

The dependence on $\xi$ and $v$ shows up starting from $V_{g^6}$. The one additional order of $g^2$ noticeably suppresses this contribution. This is shown in Figure \ref{pot3}, where we had to multiply the correction by a factor of $30$ to make it comparable with the lowest order potential. 

\begin{figure}[ht]
\begin{center}
  \includegraphics[scale=0.39]{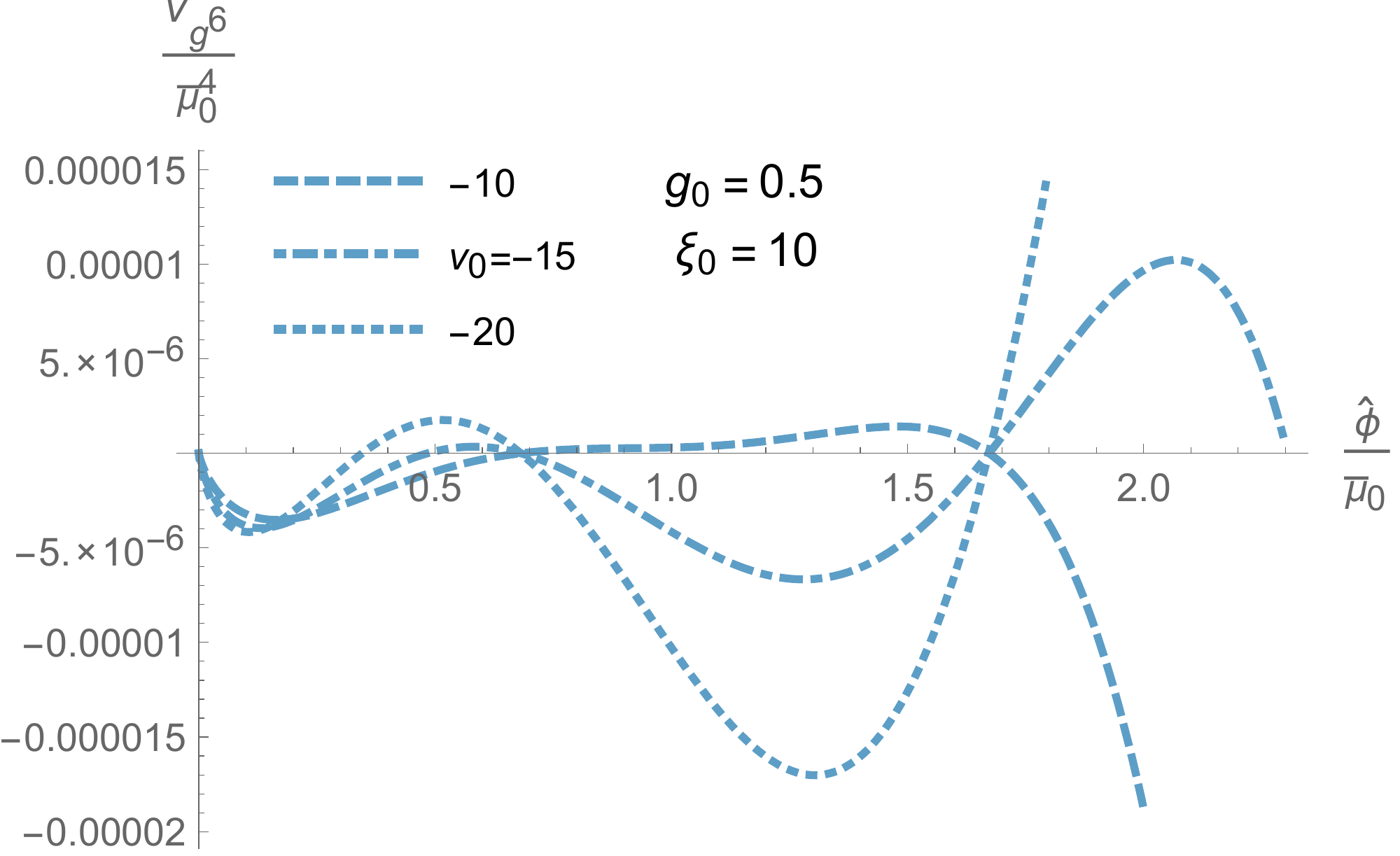}
  \includegraphics[scale=0.39]{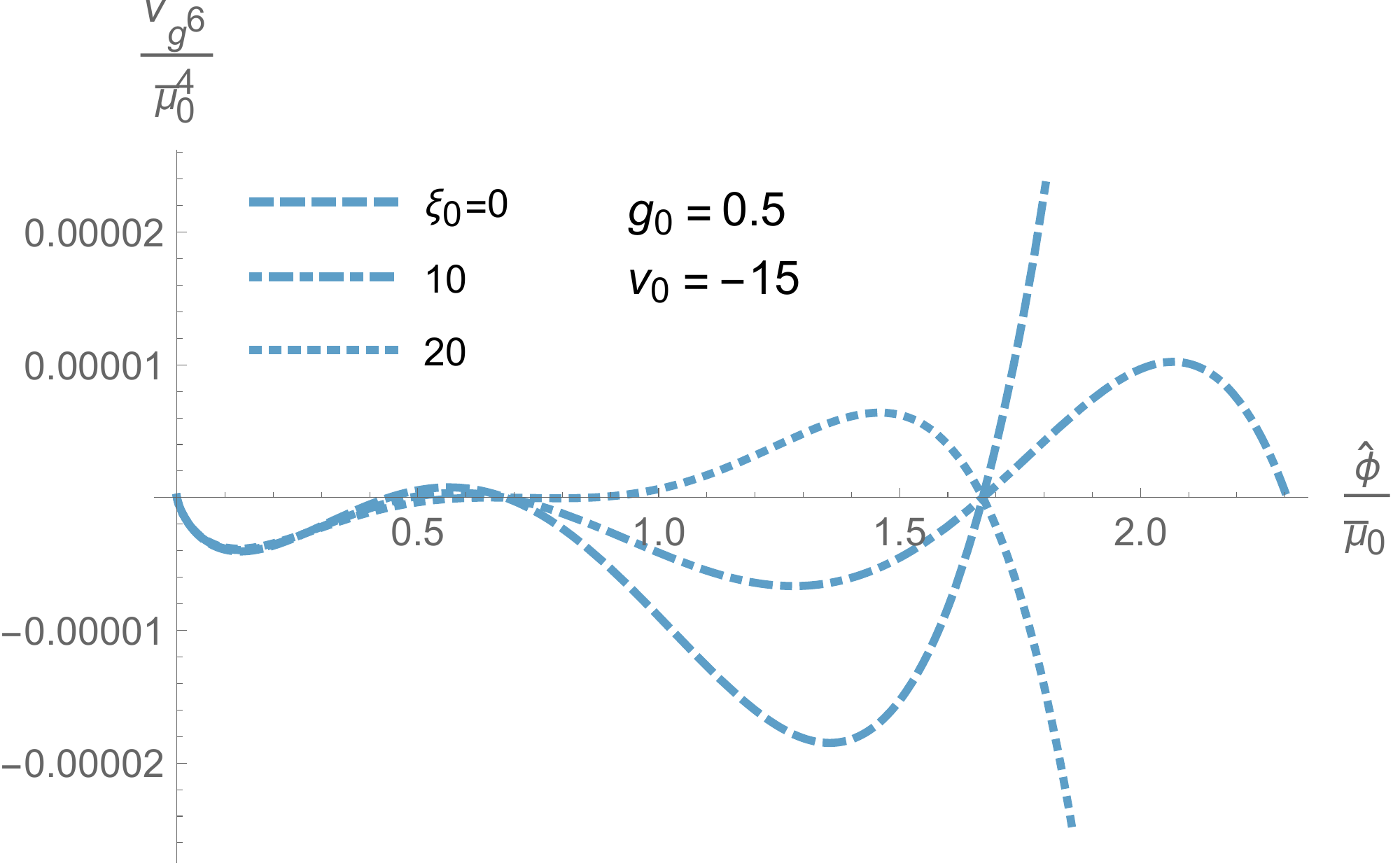}
 \end{center}
 \caption{Dependence of the potential on the gauge fixing parameters. The $v_0$ should be understood as $v_0/\bar{\mu}_0$.}
 \label{plot4}
\end{figure}

Fig. \ref{plot4} again shows plots of the $V_{g^6}$ with $c_\lambda = 1$, $c_m=1$ and $g_0=0.5$, this time for several different values of the gauge fixing parameters $\xi_0$ and $v_0$. They all cross the $x$ axis around the value $\varphi_{\text{min}} \approx 1.7$ since this is where the second minimum lies. Hence only the part of the plots to the left of this value is of interest.

\begin{figure}[ht]
\begin{center}
  \includegraphics[scale=0.4]{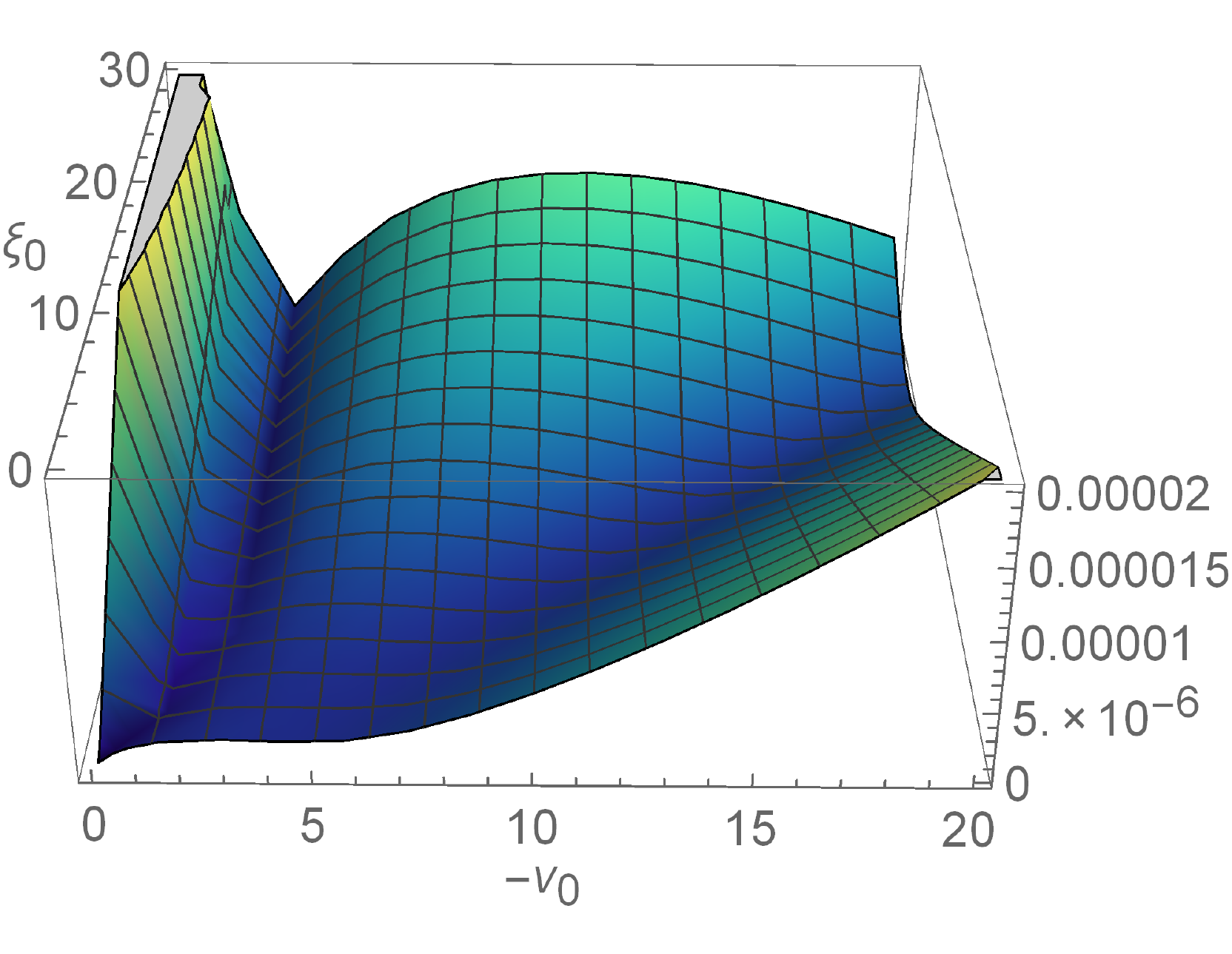}
  \includegraphics[scale=0.3]{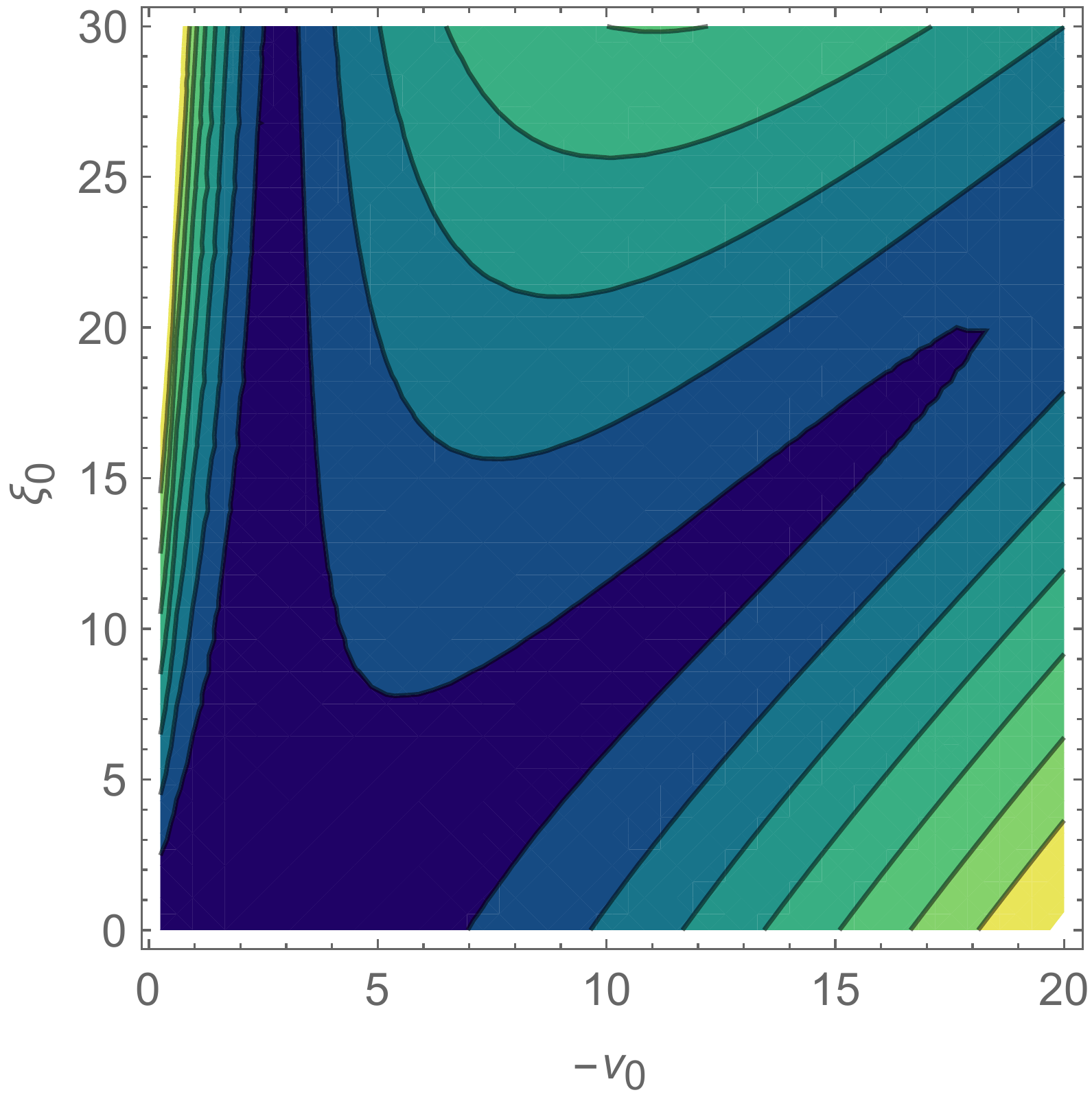}
  \includegraphics[scale=0.3]{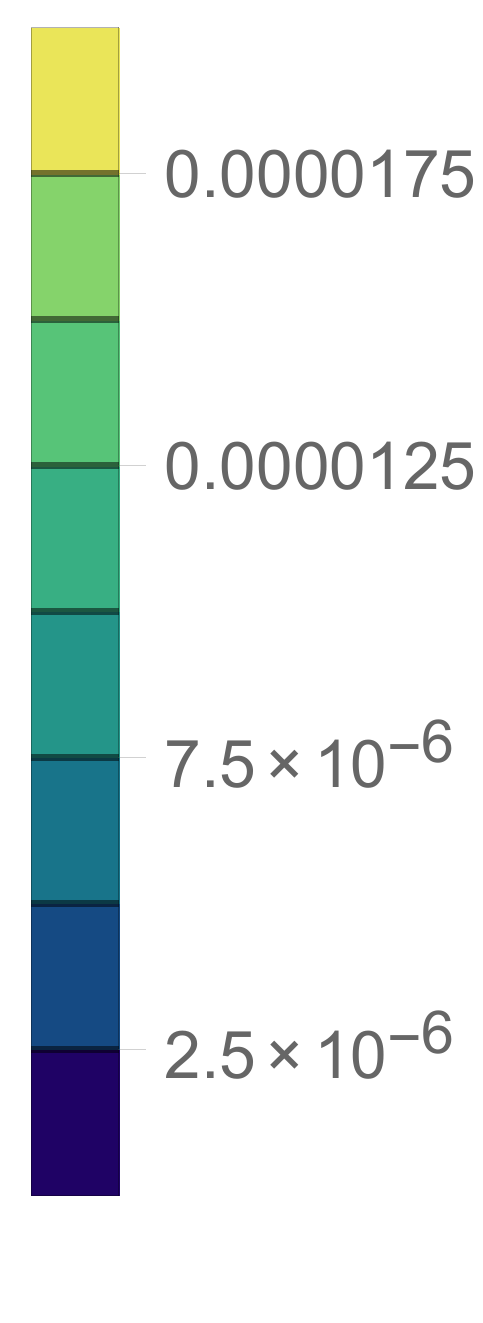}
 \end{center}
 \caption{The quantity $\Delta V$ \eqref{amount} plotted for a range of gauge fixing parameters. The $v_0$ should be understood as $v_0/\bar{\mu}_0$.}
 \label{plot3d1}
\end{figure}

The plots allow us to gain some insight into how the potential changes with the gauge fixing. One may for example notice that increasing $\xi_0$ seems to actually flatten the correction. But to quantify this fact, we define a crude measure of the overall correction,
\begin{equation}\label{amount}
\Delta V= \sqrt{\int_0^{\varphi_{\text{min}}}\left(V_{g^6}\right)^2 \; \text{d} \varphi }\;,
\end{equation}
and plot it as a function of $\xi_0$ and $v_0$ ( as before $g_0=0.5$).
The result is presented in Figure~\ref{plot3d1}.
We are reminded that the contribution blows up for $v_0 = 0$ $(\xi \neq 0)$.
But we also observe that making $-v_0$ extremely large should be met with larger values of $\xi_0$ as well, if one wishes to keep the correction as small as possible.
Also for $v_0$ around $-3$ the correction appears to be exceptionally insensitive to the value of $\xi_0$.

\subsubsection{Tunneling bounces}
We have numerically computed bounce solutions $\varphi_B$ \eqref{bouncesolution} for the potentials in Figure~\ref{potentials}. The dots show $\varphi_B^0(0)$ values of the field probed by the tunnelling instantons.
The field, having tunnelled through the barrier, ends up at the slope of the potential at $\varphi_B^0(0)$ and continues its travel by a classical roll towards the minimum. We observe a typical dependence, the further the minima are from the degenerate case, the further from the true vacuum the field emerges.

Computed values of the tree-level action $S_B^0$, as well as the correction $S_B^1$ \eqref{actions}, are summarised in Table \ref{actiontable}.
Thanks to the Nielsen identity \eqref{pertNielsen}, the obtained $S_B^1$ exibits absolutely no dependence on the gauge fixing. Even though the consistency of our perturbative calculation demands $\xi$ and $v$ to be unsuppressed by any power of $g$, we could have put $\xi_0 =0$ and $v_0 = 0$ and still obtain the $S_B^1$ given in Table \ref{actiontable}.
In this particular case the correction to the action is caused solely by the modification of the kinetic term: $K_{g^2} = 3 \frac{g_0^2}{(4 \pi^2)} \log \frac{g_0^2 \hat{\varphi}^2}{\bar{\mu}_0^2}$, see  \eqref{pertAction}.

\begin{table}[h]
\begin{center}
\begin{tabular}{c|c c}
  $g_0$ & $S_B^0$ & $S_B^1$\\
  \hline
  0.1 & 572.7 & -1.1 \\
  0.2 & 71.4 & -0.4 \\
  0.3 & 29.7 & -0.3 \\
  0.4 & 27.1 & -0.4 \\
  0.45 & 39.8 & -0.8 \\
  0.5 & 124.8 & -2.7
\end{tabular}
\caption{Action of the bounce solutions computed for the discussed potentials, see Figure~\ref{potentials}.}
\label{actiontable}
\end{center}
\end{table}
\section{Prefactor}
The semiclassical 1-loop expression for the tunneling rate (per unit volume) contains in addition to the expenent of the bounce action the prefactor. Its origin lies in the extraction of zero modes of the fluctuations around the bounce and in the functional integration over fluctuations around the bounce solution.
The well known form of the 1-loop expression is 
\begin{equation}
\frac{\gamma}{V} = \frac{S^{2}_B}{ 4 \pi^2} \sqrt{\left | \frac{{\rm \det} (-\Box_E + V''(\varphi_{min}))}{{\rm \det}' (-\Box_E + V''(\varphi_B))} \right |} e^{-S_B},
\end{equation}
where $\bar{S}_E$ is the action of the Euclidean bounce, the $\varphi_B$ denotes the bounce itself and $\varphi_{\rm f}$ is the false vacuum. The prefactor contains the square of the action of the bounce, which comes from the Jacobian of the change of variables in the path integral that replaces the integration over the 4 zero modes of the full determinant of the second variation of the action around the bounce by integration over the position of the center of the bounce.
The primed determinant is understood to have the zero-modes omitted. This expression is formally valid at 1-loop level and can easily take into account renormalisation, which is seen when one writes down the determinant as the exponent of the logarithm: 
\begin{equation} \label{decayratewpf}
\frac{\gamma}{V} = \frac{S^{2}_B}{ 4 \pi^2} e^{-\Gamma_E '[\varphi_B] + \Gamma_E[\varphi_{min}]},
\end{equation}
with 
\begin{equation}
\Gamma_E ' [\phi] = S_E [\phi] - \frac{1}{2} {\rm re}\ {\rm tr}' \log (-\Box_E + V''(\phi))\;,\quad S_B= S_E [\varphi_B] \, ,
\end{equation}
where the prime represents the omission of the zero-modes. Taking into account that the standard counterterms in the action $\Gamma_E ' [\phi]$, which cancel UV divergencies, must have the same form independently of the background, it is obvious  that they will cancel the UV divergencies encountered in the computation of the determinant. Operators with higher (than two) derivatives do not need new counterterms when computed on $\varphi_B(x)$ (see for instance \cite{ZinnJustin:2007zz}). 
 
The fact that the Jacobian contains simply the action of the bounce computed with the lowest order action is related to the canonical normalisation of the kinetic term. In fact, this happens to be the normalisation of the zero mode at the tree level in such a case. If the kinetic term in the action becomes more complicated, say $ Z(\phi) (\partial \phi)^2$, one should define the canonically normalized field according to $\widetilde{\phi}(x) = \int^{\phi(x)} \sqrt{Z(y)} dy$, which leads to a more complicated form of the scalar potential and mixes the orders of the expansion in $g$.  Generalization of the expression (\ref{decayratewpf}) to the case of higher order corrections needs some care. First of all, in the case of radiative breaking already the tree-level parameters in the potential, $\lambda$ and $m^2$, are numerically of the order $g^4$. First corrections are of the order $g^6$ in the scalar potential and order $g^2$ in the kinetic part, $Z=1+K_{g^2} (\phi)$. However, due to Nielsen identities the $g^2$ corrections are related to the $g^6$ corrections to the potential, hence it is consistent to derive the leading-order bounce configuration with the canonical kinetic term, $Z=1$, and order $g^4$ scalar potential. Since we are dealing with the stationary configuration, we shall not need to know the next correction to the bounce if we compute the tunnelling rate to the order $g^6$ only. Hence in the case under consideration it is consistent to adopt the semi-classical expression in the form 
\begin{equation}
\frac{\gamma}{V} = \frac{S^{2}_E [\phi_{b, \, g^4}]}{ 4 \pi^2} e^{-\Gamma_E [\phi_b] + \Gamma_E[\phi_f]},
\end{equation}
with 
\begin{equation}
\Gamma_E [\phi] = \int d^4x \, \left ( \frac{1}{2}\left[ 1+ K_{g^2}(\hat{\varphi}) \right]\partial_\mu \hat{\varphi} \partial^{\mu} \hat{\varphi} + (V_{g^4} + V_{g^6})(\hat{\varphi}) \right ),
\end{equation}
which is the renormalized action given in (\ref{pertAction}). 
This expression is gauge invariant. 
It  agrees with the modified perturbative expansion of the result given in \cite{Plascencia:2015pga}. In practice, given a more complicated model like the SM, one resorts to 1-loop calcultions of the determinant or to an educated parametrization of the prefactor, as discussed later in the paper. One should note, that truncation of the action at the level of second derivatives is not well justified in the case of inhomogeneous bounce background. This issue has been mentioned briefly in \cite{Lalak:2014qua} and needs further study. 

\section{Comments on gauge fixing reparametrization and RGE improvement}

There are two interesting technical questions which appear here in the context of the radiatively corrected effective action. 
We are discussing an explicit example of a model with two gauge fixing parameters, one of which is dimensionful. Gauge fixing parameters are unphysical, completely absent from observables at each and every level of a perturbative calculation. The same is true of the renormalisation scale $\mu$.
Thus we have two gauge fixing parameters which can in principle mix with each other and two dimensionful scales which parametrize radiative corrections. Hence, it is legitimate to ask the following questions. Firstly, whether arbitrary reparametrization $(\xi, v) \rightarrow (\bar{\xi},\bar{v})$  leaves  invariant the effective action for the bounce solution and secondly, how well the renormalization group improvement works in the presence of the second, in addition to $\mu$, dimensionful parameter in the action. We shall discuss these issues below in some detail. 

In the remaining part of this section we will omit the $0$ subscript in $\xi_0$ and $v_0$ for brevity.

\subsection{Reparametrizing the gauge fixing parameters}

Say we a have a new favorite pair $(\bar{\xi},\bar{v})$, so that
\begin{gather}
\begin{split} \label{repar}
\xi = \xi(\bar{\xi}, \bar{v}) \quad,\qquad
v = v(\bar{\xi}, \bar{v})\;.
\end{split}
\end{gather}
For example $\bar{\xi} = \xi$ and $\bar{v} = \xi v$.

Next we try to mimick the Nielsen identities \eqref{pertNielsen},\eqref{pertNielsen2}, but for the new parameters
\begin{gather}
\begin{split}
\frac{\partial K}{\partial \log\bar{\xi}} = \frac{\partial K}{\partial \log\xi} \frac{\partial\log\xi}{\partial\log\bar{\xi}}+ \frac{\partial K}{\partial \log v} \frac{\partial\log v}{\partial\log\bar{\xi}} = 2 \frac{\partial C^{\xi}}{\partial \varphi_1} \frac{\partial\log\xi}{\partial\log\bar{\xi}} + 2 \frac{\partial C^{v}}{\partial \varphi_1} \frac{\partial\log v}{\partial\log\bar{\xi}} = \\ = 2 \frac{\partial}{\partial \varphi_1} \left( C^{\xi} \frac{\partial\log\xi}{\partial\log\bar{\xi}} + C^{v} \frac{\partial\log v}{\partial\log\bar{\xi}} \right) =:  2 \frac{\partial}{\partial \varphi_1} C^{\bar{\xi}}
\end{split}\\
\begin{split}
\frac{\partial V_{g^6}}{\partial \log \bar{\xi}} = 
\frac{\partial V_{g^6}}{\partial \log\xi} \frac{\partial\log\xi}{\partial\log\bar{\xi}}+ \frac{\partial V_{g^6}}{\partial \log v} \frac{\partial\log v}{\partial\log\bar{\xi}} = \left( C^{\xi} \frac{\partial\log\xi}{\partial\log\bar{\xi}} + C^{v} \frac{\partial\log v}{\partial\log\bar{\xi}} \right) \frac{\partial V_{g^4}}{\partial \varphi_1} =: C^{\bar{\xi}} \frac{\partial V_{g^4}}{\partial \varphi_1}
\end{split}
\end{gather}
And similarly for derivatives wrt $v$. Thus we consistently obtain new, equally good $C$-functions
\begin{gather}\begin{split}\label{chainrule}
C^{\bar{\xi}} = C^{\xi} \frac{\partial\log\xi}{\partial\log\bar{\xi}} + C^{v} \frac{\partial\log v}{\partial\log\bar{\xi}} \\
C^{\bar{v}} = C^{\xi} \frac{\partial\log\xi}{\partial\log\bar{v}} + C^{v} \frac{\partial\log v}{\partial\log\bar{v}}
\end{split}\end{gather}
As a conclusion, the switch to $(\bar{\xi},\bar{v})$ does not spoil the Nielsen identities. Hence, the action of the bounce stays invariant under such reparametrization. 

At this point one can make an additional observation, that our $C$ functions, \eqref{Cfunctions}, satisfy
\begin{equation}\label{Csymproperty}
\frac{\partial C^v}{\partial \log\xi} = \frac{\partial C^\xi}{\partial \log v}\,,
\end{equation}
and this property is preserved by reparametrisation \eqref{repar}. We could again perform a brute force check. But this property, just as \eqref{chainrule}, follows immediately from what we already know. Namely that $C^\alpha$ is proportional to the derivative $\frac{\partial V_{g^6}}{\partial \log \alpha}$. Hence \eqref{chainrule} is simply an example of a chain rule in differentiation and \eqref{Csymproperty} expresses the symmetry of a mixed double derivative. But those properties hold irrespective of which coordinate system is used.

\subsection{RGE improvement}

Without going into a deep theoretical discussion of the implications of the presence of the second dimensionful parameter, we shall use the example worked out in this paper to see whether the correct effective action can be reconstructed with the help of the RGE improvement. 
Let us assume that we have got only the tree level Lagrangian (corrected up to the $\mathcal{O}(g^4)$ for consistency), that we wish to "RGE-improve" with higher order corrections obtained only through the renormalisation group equations. 

There is a simple contribution to RGEs at the lowest order,
\begin{equation}
\lambda(\mu)_{g^4} = \lambda + \frac{9\,g_0^4}{4 \pi^2} \log \frac{\mu}{\bar{\mu}_0} + \mathcal{O}(g^6)\;.
\end{equation}
Plugging it into the tree-level potential, we would get
\begin{equation}
V^{(R)}_{g^4} = \frac{1}{2} m^2 \hat{\varphi}^2 + \frac{1}{4!} \lambda \hat{\varphi}^4 + 3 \frac{g_0^4}{64 \pi^2} \log\! \frac{\mu^2}{\bar{\mu}_0^2}\; \hat{\varphi}^4\;.
\end{equation}
Now, assuming we also managed to compute the proper correction to the lagrangian at zeroth order, we actually have
\begin{equation}
V_{g^4} = \frac{1}{2} m^2 \hat{\varphi}^2 + \frac{1}{4!} \lambda \hat{\varphi}^4 + 3 \frac{g_0^4}{64 \pi^2} \left(\log\! \frac{g_0^2\hat{\varphi}^2}{\bar{\mu}_0^2}\,-\,\frac{5}{6}\right) \hat{\varphi}^4\;.
\end{equation}
Comparing the last two expressions, we guess that a substitution
\begin{equation}
\mu^2 \;\rightarrow\; \mu^2_A(\varphi) := g_0^2 \hat{\varphi}^2 e^{-\frac{5}{6}}
\end{equation}
could have spared us some work.

Now we set on to plug in higher order $\mu$ dependence.
Let us first take the kinetic term under consideration.
\begin{gather}
\left( 1 + K^{(R)1} \right)\left( \partial_\mu \hat{\varphi } \right)^2 = \Gamma^2 \left( \partial_\mu \varphi \right)^2 = \left[ 1 + \frac{g_0^2}{2(4 \pi)^2}(3 - \xi_0) \log \! \frac{\mu^2}{\bar{\mu}_0^2} \right]^2 \left( \partial_\mu \hat{\varphi} \right)^2 
\end{gather}
hence
\begin{gather}
K^{(R)1} \;=\; 3\frac{g_0^2}{(4 \pi)^2}\log \! \frac{\mu^2}{\bar{\mu}_0^2} \,-\, \xi_0 \frac{g_0^2}{(4 \pi)^2}\log \! \frac{\mu^2}{\bar{\mu}_0^2} \, .
\end{gather}
We could have also gone with more exotic supposition that $\mu$, newly promoted to $\varphi$, should be hit with the derivative as well:
\begin{gather}
\begin{split}
&\left( 1 + K^{(R)2} \right)\left( \partial_\mu \hat{\varphi} \right)^2 = \left[ \partial_\mu \left(\Gamma\varphi +w \right)\right]^2 =\\
&\hspace{0.5cm}= \left(\partial_\mu \hat{\varphi} \right)^2 \,+\,(3 - \xi_0) \frac{g_0^2}{(4 \pi)^2}\partial_\mu \hat{\varphi} \left( \log \! \frac{\mu^2}{\bar{\mu}_0^2}\, \partial_\mu \hat{\varphi} + \hat{\varphi} \,\partial_\mu \!\log \mu^2 \right) - 2 \frac{g_0^2}{(4 \pi)^2} v_0 \, \partial_\mu \hat{\varphi} \, \partial_\mu \log \mu^2\;.
\end{split}
\end{gather}
As we can see this second option appears as a more desirable one, since, when $\mu^2 \sim \hat{\varphi}$, it can reproduce the curious looking part of our original kinetic term of the form $\frac{v}{\varphi}$.

Let us go back to the potential,
\begin{equation}
V^{(R)}_{g^6} = - \frac{g_0^2}{2(4 \pi)^2} \log\! \frac{\mu^2}{\bar{\mu}_0^2}\,\left(2 v_0  + \xi_0 \hat{\varphi} \right)\;\frac{\partial V_{g^4}}{\partial \hat{\varphi}} \;.
\end{equation}
Unfortunately there is no way to get all the nominators under the logatithms right with just a single ansatz for $\mu^2$.
But one could go a long way towards reproducing the proper form of the correction, if he could justify a following prescription: Inside terms that vanish in the limit $
\xi_0\,,\,v_0\;\rightarrow\,0$, substitute
\begin{equation}
\mu^2 \, \rightarrow\,-g_0^2 v_0 \hat{\varphi}\,.
\end{equation}
And inside any other terms
\begin{equation}
\mu^2 \, \rightarrow\,g_0^2 \hat{\varphi}^2\,.
\end{equation}

Even so, terms build from two different logarithms, like $\log( -g_0^2 v_0 \hat{\varphi}) \log (g_0^2 \hat{\varphi}^2)$, would require carefull analysis to realise which logarithm originates from lower level correction. 

Thus we are generically unable to reconstruct the full form of the action via the RGE improvement.  

\section{Gauge independence of the vacuum lifetime in the Standard Model}
 
Now we turn to the computation of the lifetime of the SM electroweak vacuum.
Ideally one should perform a full calculation and show the gauge invariance of the result explicitly.
However the formal level gauge-independence is easily lost in the course of approximations needed in practice to obtain an analytical result.
We will discuss this point on the example of the simplest method used to estimate the lifetime of electroweak vacuum. 

We begin with a classical Lagrangian of a neutral scalar field
\begin{equation} \label{classlagr}
\Lagr = \frac{1}{2}\left( \partial \phi \right)^2 - \frac{\lambda_c}{4} \phi^4 \;,
\end{equation}
where this time $\lambda_c$ is a negative constant.
This very simple model admits an analytical bounce solution corresponding the decay of the $\phi = 0$ configuration \cite{WeinbergLee}. The corresponding action is $S=\frac{8 \pi^2}{3}\frac{1}{\lambda_c}$.
The next step is to use this solution by reinterpreting this classical Lagrangian as an effective quantum  Lagrangian of the Higgs field and simply using the well known decay rate formula which now takes the form
\begin{equation}\label{tunnelrate}
\frac{\gamma}{V} = (\text{dimensionfull quantity})^4 \,e^{-\frac{8 \pi^2}{3}\frac{1}{|\lambda_c|}}  \, .
\end{equation}
Due to the classical scale invariance present in the simple classical Lagrangian, one has no dimensionfull quantity to fill in above.
This problem can solved due to quantum corrections since even in the lowest order of perturbative calculation our actual Lagrangian reads
\begin{equation} \label{quantlagr}
\mathcal{L} = \frac{1}{2}Z(\mu)\left( \partial \phi \right)^2 - \frac{\lambda(\mu)}{4} Z(\mu)^2 \phi^4 \;.
\end{equation}
where
\begin{equation}
Z(\mu)^{\frac{1}{2}} = e^{-\int \gamma(\mu) \,\D \log(\mu)}
\end{equation}
denotes the running of the Higgs field due to a nonzero anomalous dimension, while $\lambda(\mu)$ is the running Higgs quartic coupling.
Naively using thus obtained quartic coupling in \eqref{tunnelrate} results in the following tunnelling rate
\begin{equation} \label{tunnelrate2}
\rho \sim {\Lambda^4 } e^{-\frac{8 \pi^2}{3}\frac{1}{|Z^2(\Lambda) \lambda(\Lambda) |} }\;.
\end{equation}
Now we can chose the scale $\Lambda$ to minimize the value of $ \lambda(\mu)$ which corresponds to a lower bound on the lifetime
\begin{equation}\label{lifetime}
\frac{\tau}{T_U} \sim \frac{1}{\Lambda^4 T_U^4} e^{\frac{8 \pi^2}{3}\frac{1}{|Z^2(\Lambda) \lambda(\Lambda) |} }\;.
\end{equation}
We also approximated the four-volume of our past lightcone simply by fourth power of the age of the universe which allowed us to integrate \eqref{tunnelrate} and switch from the decay rate to the lifetime.
This is a fair approximation since neglecting order one factors in front of the exponential function introduces an error much smaller than the uncertainty of the exponent.

Requiring gauge invariance of our results clearly shows a problem with this simple approximation.
While $\lambda(\mu)$ does not depend on gauge fixing the SM anomalous dimension is gauge-dependent \cite{DiLuzio:2014bua}, thus making \eqref{lifetime} very sensitive to the values of gauge fixing parameters.

To improve this result we first notice that the full coefficients in front of the quartic and kinetic terms are dimensionless, and due to the absence any dimensionfull parameters in the theory, they have to depend on $\mu$ only via $\phi/\mu$.
Secondly, running of $Z$ and $Z^2 \lambda$ fully captures the dependence of these coefficients on $\mu$, which means that it cancels between explicit dependence in loop corrections and running of the couplings.
In conclusion, to improve the accuracy of our approximation at any field value we can replace the scale with the value of the field, $\mu \rightarrow \phi$, and  the resulting $\mu$-independent function of $\phi$ will be a good approximation of the full effective quantum Lagrangian. 
This leads to
\begin{equation} \label{quantlagr2}
\Lagr = \frac{1}{2}\left( \partial Z^{\frac{1}{2}}(\phi) \phi \right)^2 - \frac{\lambda(\phi)}{4} Z^2(\phi) \phi^4 \;.
\end{equation}
The above Lagrangian suggests that it makes sense to redefine the field variable by $\tilde{\phi} = Z^{\frac{1}{2}}(\phi) \phi$, which completely eliminates the field renormalisation $Z(\mu)$.
This brings us to the point: the $|Z^2 \lambda|$ in  \eqref{lifetime} should simply be replaced by $|\lambda|$ alone,
\begin{equation}\label{lifetime2}
\frac{\tau}{T_U} \sim \frac{1}{\Lambda^4 T_U^4} e^{\frac{8 \pi^2}{3}\frac{1}{| \lambda(\Lambda) |} }\;,
\end{equation}
where the absence of $Z$ trivially makes the result gauge-independent.
This approach is connected with the treatment of abelian gauge theory we discussed in the previous sections, where it was shown that for formal consistency, including quantum corrections to the kinetic term is crucial. 
It is important to stress that replacing $\mu$ with $\phi$ requires treating also $\phi$ in $Z(\phi)$, as a spacetime dependent configuration, and thus hitting $Z(\phi(x))$ with the derivative $\partial_\mu$.

\begin{figure}[ht]
\centering
\includegraphics[scale=1.0]{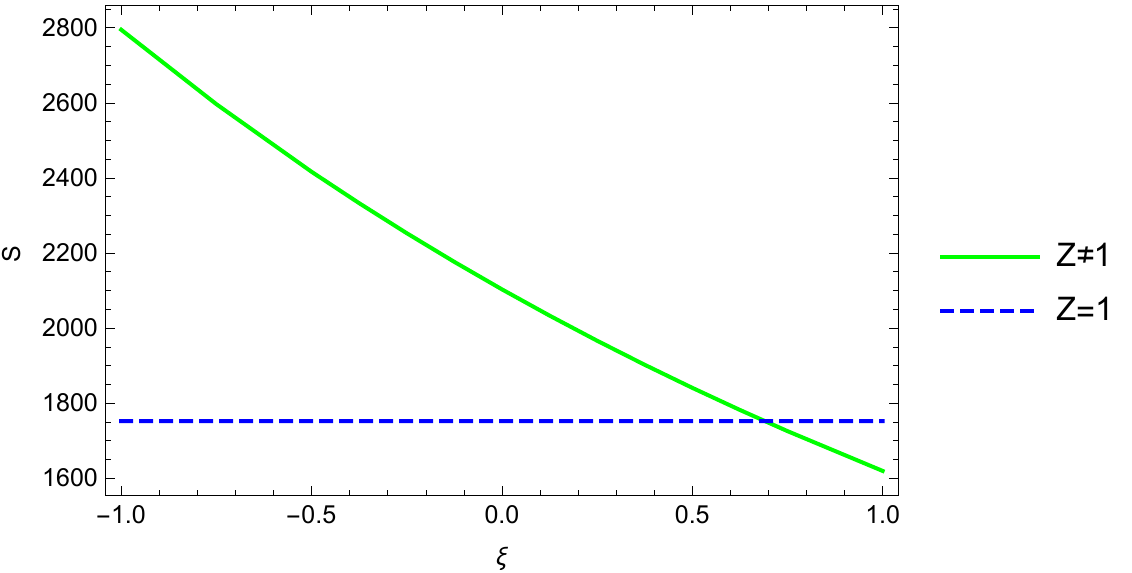} 
\caption{Solid line: gauge dependence of the bounce action calculated including the field renormalisation factor $Z$ only in the potential. Dashed line: the action calculated after the redefinition $\tilde{\phi} = Z^\frac{1}{2}(\phi) \phi$. $\xi$ is a gauge fixing parameter.} 
\label{SodXi}
\end{figure}

In the Standard Model, difference between \eqref{lifetime} and \eqref{lifetime2} is very significant.
The $Z^2(\mu)$ changes from $1.0$ at the scale of the top mass, $M_{top}$, to about $0.8$ close to the Planck mass, when computed in Landau gauge.
This dependence ends up in the exponent of \eqref{lifetime} increasing it from roughly $1800$ to $2100$. 
As a result the lifetime of the electroweak vacuum compared to the lifetime of the universe, computed via \eqref{lifetime} is around $10^{676}$ while properly using \eqref{lifetime2} gives $10^{529}$. Figure \ref{SodXi} illustrates gauge dependence of the action, using the simplest class of gauge fixing, the so called Fermi gauges, with $\xi = \xi_W(M_{top})=\xi_B(M_{top})$.
Landau gauge is an example of this class as an RGE-stable choice of $\xi =0$. Thus incorrectly including the field renormalisation in the SM, even in Landau gauge results in a large overestimation of the expected lifetime.

Now we turn to the possible nonrenormalizable terms in scalar potential, e.g. 
$
\frac{\lambda_6}{6!\, M^2} \phi^6\;.
$
In our simplified approach the effective contribution of this term reads
\begin{equation}
\frac{\lambda_6(\phi)}{6!\, M^2} Z^3(\phi) \phi^6\;,
\end{equation}
with $\lambda_6$ contributing to RGE's of other couplings.
However, it does not appear in $\gamma$ at one loop. Again, since running of $\lambda_6(\mu)$ does not depend on the gauge fixing, absorbing the field renormalisation into redefinition of the field, we end up without any gauge dependence.

There is another important point to be made about gauge dependence of nonrenormalizable terms in the potential, which was detailed in earlier sections in the case of the simple abelian model.
Nielsen identities bind variation of the effective action to its derivative with respect to gauge parameters \cite{Nielsen:1975fs}. Taking momentum independent part of this relation (which means going from effective action to effective potential), one arrives at
\begin{equation}\label{Nielsen}
\xi \frac{\partial V(\phi)}{\partial \xi} = C^\xi (\phi) \frac{\partial V(\phi)}{\partial \phi}\;.
\end{equation}
The function $C$ admits a perturbative expansion, like other parts of the equation \eqref{Nielsen}. However then it inherently involves ghost, goldstone and gauge propagators \cite{Aitchison}. As a result vertices produced by nonrenormalizable operators, like the six-legged $\lambda_6$, do not contribute to $C$ at the lowest levels of the expansion.
 When we classify terms on the left and right hand side of \eqref{Nielsen} by their power of $M$, we see immediately that the same $C$ function separately governs gauge dependence of both renormalizable and nonrenormalizable part of the potential with any dimension.
As a result, for practical purposes, it is enough to uncover gauge dependence of the renormalizable part of the Lagrangian, to be able fully reconstruct it for all operators including those of higher dimension.

\begin{figure}[ht]
\centering
\includegraphics[scale=1.0]{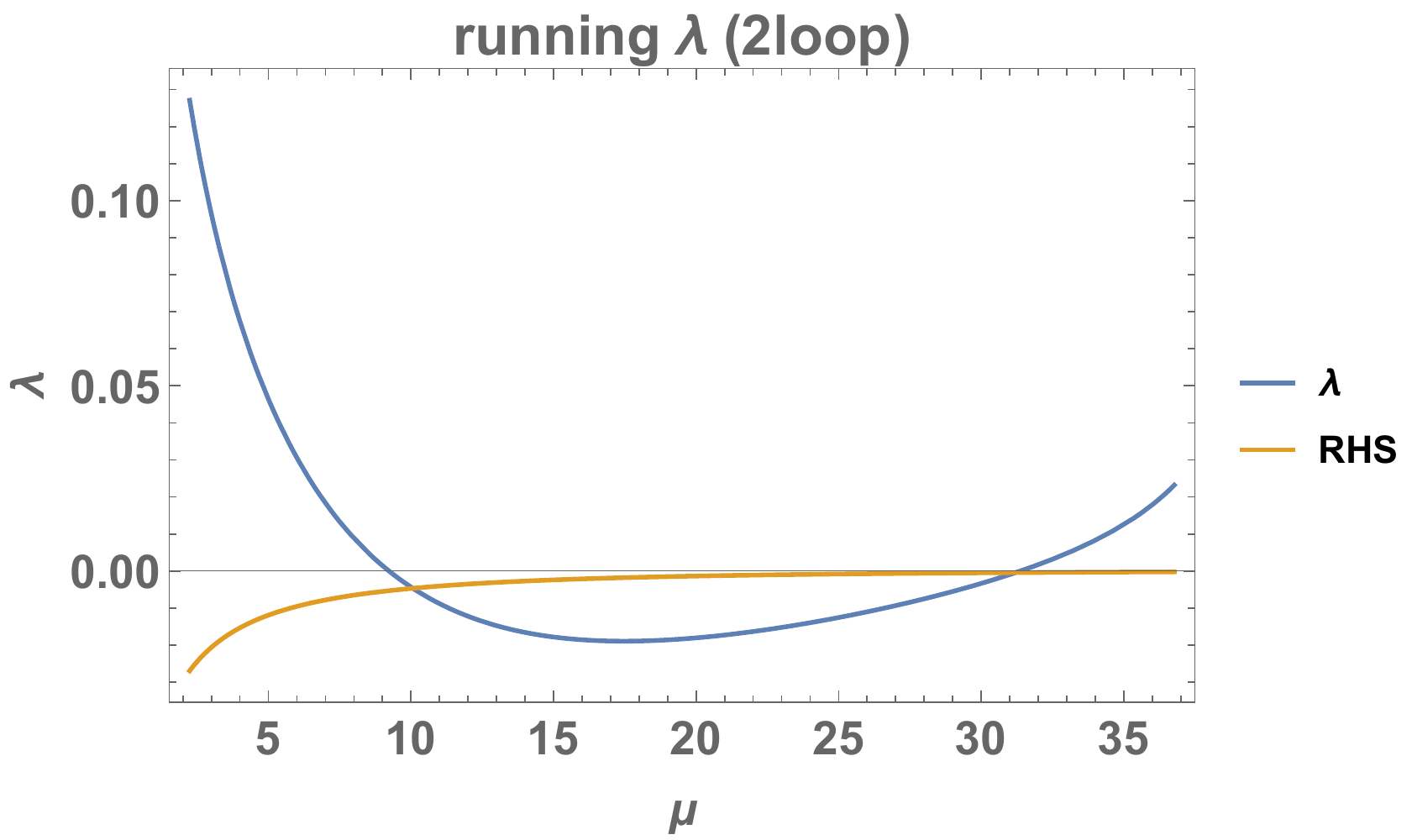} 
\caption{Running of the quartic coupling $\lambda$ vs running of the right hand side of the formula (\ref{RHS}). The points where the curves meet mark the radiatively generated extrema of the SM effective potential - first from the left is the maximum and the second is the stable minimum. In general the relation between couplings which holds at the extrema becomes violated, and the region where the violation is significant contributes to the action of the bounce.} 
\label{LvsRHS}
\end{figure}

Finally, let us pause to comment on the validity of the "radiative" expansion based on the relation $\lambda \sim g^4$ in case of the SM. 
The relation between couplings which holds in the SM at the 1-loop order at the critical points of the effective potential reads
\begin{equation} \label{RHS}
\lambda = \frac{\hbar}{256 \pi^2} \left [ g^{4}_1 + 2 g^{2}_1 g^{2}_2 + 3 g^{4}_2 - 48 h^{4}_{t}  - 3 (g^{2}_1 + g^{2}_{2} )^2  \log \frac{g^{2}_1 + g^{2}_{2}}{4 } -
6 g^{4}_2 \log \frac{g^{2}_{2}}{4} + 48 y^{4}_{t} \log \frac{y^{2}_t}{2} \right ] .
\end{equation}
The running of the quartic coupling $\lambda$ vs running of the right hand side of the formula \eqref{RHS} is shown in Figure~\ref{LvsRHS}.
The points where the curves meet, mark the radiatively generated extrema of the effective potential. The first from the left is the maximum and the second is the global minimum.
In general this relation between couplings which holds at the extrema is violated, and the region where the violation is significant also contributes to the action of the bounce. 
Additionally, for the inhomogenous bounce solution the momentum expansion becomes problematic and the real role of the prefactor is somewhat obscure, since some corrections are already taken into account in the effective action. 

Perhaps playing with numerical values of various contributions could lead to identification of the best approximation to the complete expression, but in practice considerations based on global measures like the one given in the expression (\ref{amount}) could be useful. 

\section{Summary}
In this paper we have investigated at the perturbative level the gauge fixing independence of the tunneling rate to a stable radiatively induced vacuum in the abelian Higgs model  in a class of $R_\xi$ gauges, in the presence of both dimensionless and dimensionful gauge fixing parameters. We performed explicit calculations in the spirit of improved perturbative expansion which assumes the quartic coupling, and the tree-level mass parameter, to be of the order of the fourth power of the gauge coupling.
We also explicitly showed gauge fixing independence of the tunnelling rate, depth of the extrema and the value of the physical pole mass. 
We also proved that Nielsen identities survive the inclusion of higher order operators.
We also discussed the applicability of these results to the Standard Model.
Unfortunately, finding the appropriate improved expansion scheme in the SM doesn't seem to be practically feasible at the next to leading order of radiative corrections, but it is important to understand reasons for the presence of the gauge-fixing non-invariance of numerical calculations and this is the aim of the present paper. We have discussed the independence of the bounce action with respect to reparametrisation of the gauge fixing. The presence of the dimensionful gauge fixing parameter introduces a second mass scale into the RG running, which makes the RGE improvement procedure for the effective action less useful. 

\pagebreak  
\begin{center}
{\bf Acknowledgements}
\end{center} 
ZL thanks DESY Theory Group for hospitality. This work was supported by the German Science Foundation (DFG) within the Collaborative Research Center (SFB) 676 "Particles, Strings and the Early Universe".
This work has been supported by National Science Centre under research grants DEC-2012/04/A/ST2/00099 and 
2014/13/N/ST2/02712. ML was supported  by doctoral scholarship number 2015/16/T/ST2/00527.

\clearpage
\section*{Appendix} \label{App1}
\appendix

\begin{table}[h!]
\renewcommand{\arraystretch}{4}
\begin{tabular}{|@{\hskip 0.6cm}c@{\hskip 1.6cm} c@{\hskip 0.3cm} l@{\hskip 0.6cm}|}
  \hline
\multicolumn{3}{|c|}{
\parbox[c][11em][c]{38em}{\parbox[c][1em][c]{7.5em}{\includegraphics[scale=0.72]{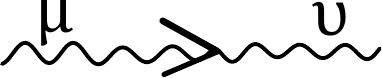}} {\Large$\displaystyle=\frac{-i}{k^2-g^2 {\varphi^\circ}^2}\left( g^{\mu \nu}- \frac{k^\mu k^\nu}{k^2}\right)\,+$\vspace{0.3cm}} \\ \parbox[c][1em][c]{17em}{\hspace{0cm}}\Large$\displaystyle +\,\frac{-i\xi (k^2 \!-\! m^2 \!-\! \frac{\lambda}{6}{\varphi^\circ}^2 )+i g^2 v^2}{D_N}\left(\frac{k^\mu k^\nu}{k^2}\right) $ }
}\\[-0.5cm]
\parbox[c][4em][c]{9em}{\includegraphics[scale=0.7]{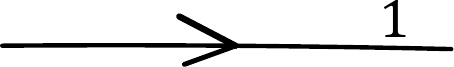}} {\Large$\displaystyle =\;\frac{i}{k^2-m^2-\frac{\lambda}{2} {\varphi^\circ}^2} $} &
\parbox[c][4em][c]{5.5em}{\includegraphics[scale=0.72]{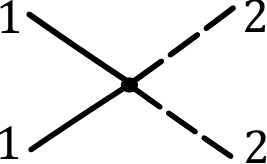}} & {\Large$\displaystyle=\;-i\,\frac{\lambda}{3}$}\\[0.4cm]
\parbox[c][4em][c]{9em}{\includegraphics[scale=0.72]{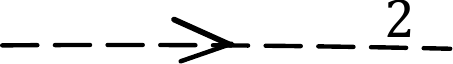}} {\Large$\displaystyle =\;\frac{i(k^2 - \xi g^2 {\varphi^\circ}^2)}{D_N} $} &
\parbox[c][4em][c]{5.5em}{\includegraphics[scale=0.72]{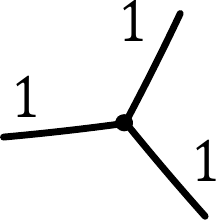}} & {\Large$\displaystyle=\;-i\,\lambda \varphi^\circ$} \\[0.4cm]
\parbox[c][4em][c]{9em}{\includegraphics[scale=0.72]{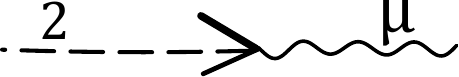}}  {\Large$\displaystyle=\;\frac{g(\xi \varphi^\circ + v)k^\mu}{D_N} $} &
\parbox[c][4em][c]{5.5em}{\includegraphics[scale=0.72]{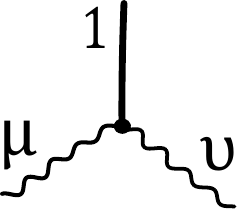}} & {\Large$\displaystyle=\;2i\,g^2 \varphi^\circ g^{\mu\nu}$} \\[0.6cm]
\parbox[c][4em][c]{9em}{\includegraphics[scale=0.72]{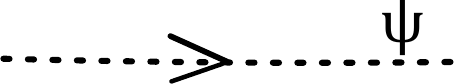}}  {\Large$\displaystyle=\;\frac{i}{k^2+g^2 v \varphi^\circ} $} &
\parbox[c][4em][c]{5.5em}{\includegraphics[scale=0.72]{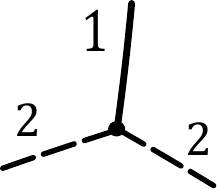}} & {\Large$\displaystyle=\;-i\,\frac{\lambda}{3} \varphi^\circ$} \\[0.8cm]
\parbox[c][4em][c]{4em}{\includegraphics[scale=0.72]{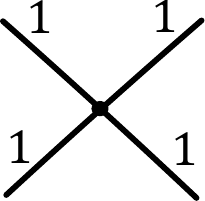}} =\hspace{0.3cm}\parbox[c][4em][c]{4.3em}{\includegraphics[scale=0.72]{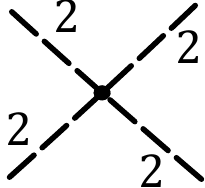}} {\Large$\displaystyle=\;-i\,\lambda$} &
\parbox[c][4em][c]{5.5em}{\includegraphics[scale=0.72]{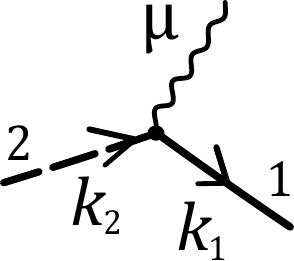}} & {\Large$\displaystyle=\;g(k_1+k_2)^\mu$} \\[1cm]
\parbox[c][4em][c]{5.5em}{\includegraphics[scale=0.72]{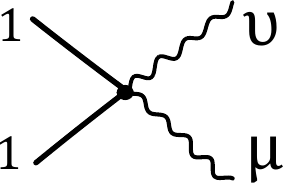}} =\hspace{0.3cm}\parbox[c][4em][c]{5.5em}{\includegraphics[scale=0.72]{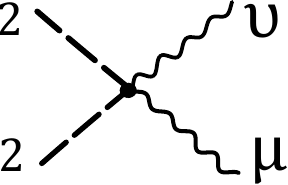}} {\Large$\displaystyle=\;2i\,g^2 g^{\mu\nu}$} &
\parbox[c][4em][c]{5.5em}{\includegraphics[scale=0.72]{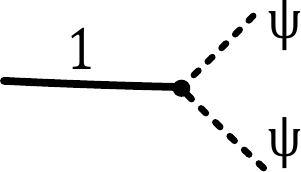}} & {\Large$\displaystyle=\;ig^2v\quad$} \\[0.6cm]
  \hline
\end{tabular} 
\caption{Feynman rules in the studied model with background field value $\varphi^\circ$ and gauge fixing parameters $\xi$ and $v$; $\displaystyle D_N = k^4 - k^2(m^2 + \frac{\lambda}{6} {\varphi^\circ}^2 - 2 g^2 \varphi^\circ v) + g^2{\varphi^\circ}^2\left[\xi\left(m^2 + \frac{\lambda}{6} {\varphi^\circ}^2 \right) + g^2 v^2\right]$.}
\end{table}

\end{document}